\documentclass[11pt,reqno]{amsart}

\usepackage{amssymb, amsthm, amsmath}

\def\lanbox{\hbox{$\, \vrule height 0.25cm width 0.25cm depth 0.01cm \,$}}
\renewcommand{\Box}{\lanbox}

\def\uprho{\raise1pt\hbox{$\rho$}}

\def\R{{\mathbb R}}

\def\mfr#1/#2{\hbox{${\frac{#1}{#2}}$}}

\newcommand{\brtf}{\bar\rho}
\newcommand{\Egp}{E^{\rm GP}}

\def\R{{\mathbb R}}

\def\E{{\mathcal E}}

\def\x{{\vec x}}
\def\p{{\vec p}}
\def\z{{\vec z}}
\def\q{{\vec q}}
\def\y{{\vec y}}
\def\k{{\vec k}}
\def\0{{\vec 0}}
\def\vecmu{{\vec \mu}}
\def\vecnu{{\vec \nu}}

\def\mfr#1/#2{\hbox{$\frac{{#1}}{{#2}}$}}
\def\uprho{\raise1pt\hbox{$\rho$}}
\def\upchi{\raise1pt\hbox{$\chi$}}
\def\dlambda{\lower1pt\hbox{$\lambda$}}
\newcommand{\xij}{|\x_i-\x_j|}

\newcommand{\rtf}{\rho^{\rm TF}}

\newcommand{\mtf}{\mu^{\rm TF}}

\newcommand{\hw}{{\mathord{\widehat{w}}^{\phantom{*}}}}
\newcommand{\hn}{{\mathord{\widehat{n}}}}
\newcommand{\an}{{\mathord{a}}^{\phantom{*}}}
\newcommand{\bn}{{\mathord{b}}^{\phantom{*}}}
\newcommand{\cA}{{\mathord{\mathcal A}}}
\newcommand{\cB}{{\mathord{\mathcal B}}}

\newtheorem{theorem}{Theorem}[section]
\newtheorem{lemma}[theorem]{Lemma}
\newtheorem{corollary}[theorem]{Corollary}

\theoremstyle{definition}

\theoremstyle{remark}

\numberwithin{equation}{section}

\newcommand{\eps}{\varepsilon}
\newcommand{\K}{{\mathcal K}}
\newcommand{\X}{{\vec X}}
\newcommand{\Tr}{{\rm Tr}\, }
\newcommand{\half}{\mbox{$\frac{1}{2}$}}

\newcommand{\pgp}{\phi^{\rm GP}}
\newcommand{\al}{\alpha}
\newcommand{\rmax}{\rho_{\al,{\rm max}}}
\newcommand{\rmin}{\rho_{\al,{\rm min}}}
\newcommand{\g}{g}

\begin{document}

\title{THE GROUND STATE OF THE BOSE GAS}

\author[E.H.~Lieb]{ELLIOTT H.\ LIEB}
\address [E.H.~Lieb] {Departments  of Mathematics and Physics, Princeton
University, Jadwin Hall,  P.O. Box 708, Princeton, NJ  08544, USA}
\email{lieb@math.princeton.edu}

\author[R.~Seiringer]{ROBERT SEIRINGER} 
\address[R.~Seiringer]{Departments  of Mathematics and Physics, Princeton
University, Jadwin Hall,  P.O. Box 708, Princeton, NJ  08544, USA. On
leave of absence from Institut f\"ur Theoretische Physik, Universit\"at
Wien, Boltzmann\-gasse 5, A 1090 Vienna, Austria}
\email{rseiring@math.princeton.edu}   

\author[J.P.~Solovej]{JAN PHILIP SOLOVEJ}
\address[J.P.~Solovej]{University of Copenhagen, Universitetsparken 5,
DK-2100 Copenhagen, Denmark}
\email {solovej@math.ku.dk}

\author[J.~Yngvason]{JAKOB~YNGVASON}
\address[J.~Yngvason]{Institut f\"ur Theoretische Physik, Universit\"at
Wien, Boltzmanngasse 5, A 1090 Vienna, Austria}
\email{yngvason@thor.thp.univie.ac.at}

\thanks{\copyright 2002 by the authors. Reproduction of this article, in its
entirety, by any means, is permitted for non-commercial purposes.}

\thanks{This article appears in {\it Current Developments in Mathematics, 2001},
International Press, Cambridge, 2002, pp. 131--178.}

\dedicatory{  }

\begin{abstract}
Now that the low temperature properties of quantum-\-mech\-anical many-body
systems (bosons) at low density, $\rho$, can be examined experimentally
it is appropriate to revisit some of the formulas deduced by many authors
4-5 decades ago.  For systems with repulsive (i.e.\ positive) 
interaction potentials the
experimental low temperature state and the ground state are effectively
synonymous -- and this fact is used in all modeling. In such cases, the
leading term in the energy/particle is $2\pi\hbar^2 a \rho/m$ where $a$ is 
the
scattering length of the two-body potential.  Owing to the delicate and
peculiar nature of bosonic correlations (such as the strange $N^{7/5}$
law for charged bosons), four decades of research failed to establish this
plausible formula rigorously. The only previous lower bound for the energy
was found by Dyson in 1957, but it was 14 times too small. The correct
asymptotic formula has recently been obtained by us and this work will
be presented. The reason behind the mathematical difficulties will be
emphasized. A different formula, postulated as late as 1971 by Schick,
holds in two-dimensions and this, too, will be shown to be correct.
With the aid of the methodology developed to prove the lower bound for the
homogeneous gas, two other problems have been successfully addressed. One
is the proof by us that the Gross-Pitaevskii equation correctly describes
the ground state in the `traps' actually used in the experiments. For
this system it is also possible to prove complete Bose condensation,
as we have shown.  Another topic  is a proof that Foldy's 1961 theory
of a high density Bose gas of charged particles correctly describes
its ground state energy. All of this is quite recent work and it is
hoped that the mathematical methodology might be useful, ultimately,
to solve more complex problems connected with these interesting systems.
\end{abstract}
\maketitle

\section*{Foreword}
At the conference ``Contemporary Developments in Mathematics'', hosted
by the MIT and Harvard University Mathematics Departments, November
16--17, 2001, one of us (E.H.L.) contributed a talk with the title
``The Bose gas: A subtle many-body problem''. This talk covered
material by all the authors listed above. This contribution is a much
expanded version of the talk and of \cite{L3}.

\tableofcontents

\section{Introduction}\label{intro}

Schr\"odinger's equation of 1926 defined a new mechanics whose
Hamiltonian is based on classical mechanics, but whose
consequences are sometimes non-intuitive from the classical point
of view. One of the most extreme cases is the behavior of the
ground (= lowest energy) state of a many-body system of particles.
Since the ground state function $\Psi(\x_1,...,\x_N)$ is
automatically symmetric in the coordinates $\x_j\in \R^3$ of the
$N$ particles, we are dealing necessarily with {\it `bosons'.} If
we imposed the Pauli exclusion~principle (antisymmetry) instead,
appropriate for electrons,  the outcome would look much more
natural and, oddly, more classical. Indeed, the Pauli principle is
{\it essential} for understanding the stability of the ordinary
matter  that surrounds us.

Recent experiments have confirmed some of the bizarre properties of
bosons close to their ground state, but the theoretical ideas go back to
the 1940's -- 1960's. The first sophisticated analysis of a gas or liquid
of {\it interacting} bosons is due to Bogolubov in 1947. His
approximate theory as amplified by others, is supposed to be exact in
certain limiting cases, and some of those cases have now been verified
rigorously (for the ground state energy) --- 3 or 4 decades after they were
proposed.

The discussion will center around four main topics.

\begin{enumerate}
\item
{ The dilute, homogeneous Bose gas with repulsive interaction (2D and 3D)}.
\item
{ Repulsive bosons in a trap (as used in recent
experiments) and the `Gross-Pitaevskii\rq\ equation}.
\item{ Bose-Einstein condensation for dilute trapped
gases}.
\item{ Foldy's `jellium' model of charged particles
in a neutralizing background}.
\end{enumerate}

Note that for potentials that tend to zero at infinity `repulsive' and
`positive' are synonymous --- in the quantum mechanical literature at
least.  In classical mechanics, in contrast, a potential that is
positive but not monotonically decreasing is not called repulsive.

The discussion below of topic 1 is based on \cite{LY1998} and
\cite{LY2d}, and of topic 2 on \cite{LSY1999} and \cite{LSY2d}. See
also
\cite{LYbham, LSYdoeb, S4, LSYnn}. The discussion of
topic 3 is mainly taken from
\cite{LS02}, but for transparency 
we also include here a section on the special case when
the trap is a rectangular box. This case already contains the 
salient points, but avoids several complications due the the inhomogeneity of
the gas in a general trap.
The discussion of topic 4 is based on \cite{LS}. 

Topic 1 (3-dimensions) was the starting point and contains essential
ideas. It is explained here in some detail and is taken, with minor
modifications (and corrections), from \cite{LYbham}. In terms of
technical complexity, however, the fourth topic is the most involved
and can not be treated here in full detail.

The interaction potential between pairs of particles in the Jellium
model in topic 4 is the repulsive, {\it long-range} Coulomb potential,
while in topics 1--3 it is assumed to be repulsive and {\it short
range}. For alkali atoms in the recent experiments on Bose Einstein
condensation the interaction potential has a repulsive hard core, but
also a quite deep attractive contribution of van der Waals type and
there are many two body bound states \cite{PS}. The Bose condensate
seen in the experiments is thus not the true ground state (which would
be a solid) but a metastable state. Nevertheless, it is usual to model
this metastable state as the ground state of a system with a repulsive
two body potential having the same scattering length as the true
potential, and this is what we shall do. In this paper all potentials
will be positive.

\bigskip

\section{The Dilute Bose Gas in 3D} \label{sect3d}

We consider the Hamiltonian for $N$ bosons of mass
$m$ enclosed in a cubic box $\Lambda$ of side length $L$ and interacting by a
spherically symmetric pair potential
$v(|\x_i - \x_j|)$:
\begin{equation}\label{ham}
H_{N} = - \mu\sum_{i=1}^{N} \Delta_i +
\sum_{1 \leq i < j \leq N} v(|\x_i - \x_j|).
\end{equation}
Here  $\x_i\in\R^3$, $i=1,\dots,N$ are the positions of the
particles, $\Delta_i$ the Laplacian with respect to $\x_{i}$, and
we have denoted ${\hbar^2}/{ 2m}$ by $\mu$ for short. (By choosing
suitable units $\mu$ could, of course, be eliminated, but we want
to keep track of the dependence of the energy on  Planck's
constant and the mass.) The Hamiltonian (\ref{ham}) operates on
{\it symmetric} wave functions in $L^2(\Lambda^{N}, d\x_1\cdots
d\x_N)$ as is appropriate for bosons. The interaction potential
will be assumed to be {\it nonnegative} and to decrease faster
than $1/r^3$ at infinity.

We are interested in the ground state energy $E_{0}(N,L)$ of (\ref{ham}) in
the
{\it thermodynamic limit} when $N$ and $L$ tend to infinity with the
density $\rho=N/L^3$ fixed. The energy per particle in this limit is
\begin{equation}\label{eq:thmlimit} e_{0}(\rho)=\lim_{L\to\infty}E_{0}(\rho L^3,L)/(\rho
L^3).\end{equation}
Our results about $e_{0}(\rho)$ are based on estimates on
$E_{0}(N,L)$
for finite $N$ and $L$, which are important, e.g., for the considerations of
inhomogeneous systems in \cite{LSY1999}.
To define  $E_{0}(N,L)$ precisely one
must specify the boundary conditions. These should not matter for the
thermodynamic limit.
To be on the safe side we use Neumann boundary conditions for the
lower bound, and Dirichlet boundary conditions for the upper bound
since these lead, respectively, to the lowest and the highest energies.

For experiments with dilute gases the {\it low density asymptotics} of
$e_{0}(\rho)$ is of importance. Low density means here that the mean
interparticle distance, $\rho^{-1/3}$ is much larger than the
{\it scattering length} $a$ of the potential,
which is defined as follows. The zero energy scattering
Schr\"odinger equation
\begin{equation}\label{3dscatteq}
-2\mu \Delta \psi + v(r) \psi =0
\end{equation}
has a solution  of the form, asymptotically as $|\x|=r\to \infty$
(or for all $r>R_0$ if $v(r)=0$ for $ r>R_0$),
\begin{equation}\label{3dscattlength}
\psi_0(\x) = 1-a/|\x|
\end{equation}
(The factor $2$ in (\ref{3dscatteq}) comes from the reduced mass of the
two particle problem.) Writing $\psi_0(\x)=u_0(|\x|)/|\x|$ this is the
same as
\begin{equation}a=\lim_{r\to\infty}r-\frac{u_{0}(r)}{u_{0}'(r)},\end{equation}
where $u_{0}$ solves the zero energy (radial) scattering equation,
\begin{equation}\label{scatteq}
-2\mu u_{0}^{\prime\prime}(r)+ v(r) u_{0}(r)=0
\end{equation}
with $u_{0}(0)=0$. 

An important special case is the hard core potential $v(r)= \infty$ if
$r<a$ and  $v(r)= 0$ otherwise. Then the scattering length $a$ and the
radius $a$ are the same.

Our main result is a
rigorous proof of the formula
\begin{equation} e_{0}(\rho)\approx4\pi\mu\rho a\end{equation}
for $\rho a^3\ll 1$, more precisely of
\begin{theorem}[\textbf{Low density limit of the ground state
energy}]\label{3dhomthm}
\begin{equation}\label{basic}
\lim_{\rho a^3\to 0}\frac {e_{0}(\rho)}{4\pi\mu\rho a}=1.
\end{equation}
\end{theorem}
This formula is independent of the boundary conditions used for the
definition of $e_{0}(\rho)$ . It holds for every positive radially 
symmetric pair 
potential such that $\int_R^\infty v(r)r^2 dr<\infty$ for some $R$, which 
guarantees a finite scattering length, cf.\ Appendix A in \cite{LY2d}.

The genesis of an understanding of $e_{0}(\rho)$ was the pioneering
work \cite{BO} of Bogolubov, and in the 50's and early 60's several
derivations of (\ref{basic}) were presented \cite{Lee-Huang-YangEtc},
\cite{Lieb63}, even including higher order terms:
\begin{equation}\frac{e_{0}(\rho)}{4\pi\mu\rho a}=
1+\mfr{128}/{15\sqrt \pi}(\rho a^3)^{1/2}
+8\left(\mfr{4\pi}/{3}-\sqrt 3\right)(\rho a^3)\log (\rho a^3)
+O(\rho a^3)
\end{equation}
These early developments are reviewed in \cite{EL2}. They all rely
on some special assumptions about the ground state that have never been
proved, or on the selection of special terms from a perturbation series
which likely diverges. The only rigorous estimates of this period were
established by Dyson, who derived the following bounds in 1957 for a
gas of hard spheres \cite{dyson}:
\begin{equation} \frac1{10\sqrt 2} \leq
    \frac{e_{0}(\rho)}{ 4\pi\mu\rho a}\leq\frac{1+2 Y^{1/3}}{
(1-Y^{1/3})^2}
\end{equation}
with $Y=4\pi\rho a^3/3$. While the upper bound has the asymptotically
correct form, the lower bound is off the mark by a factor of about 1/14.
But for about 40 years this was the best lower bound available!

Under the assumption that (\ref{basic}) is a correct asymptotic
formula for the energy, we see at once that understanding it
physically, much less proving it, is not a simple matter.
Initially, the problem presents us with two lengths, $a \ll
\rho^{-1/3}$ at low density. However, (\ref{basic}) presents us
with another length generated by the solution to the problem. This
length is the de Broglie wavelength, or  `uncertainty principle' length 
(sometimes called `healing length')
\begin{equation}\label{ellc}\ell_c\sim (\rho a)^{-1/2} . \end{equation}
 The reason for saying that
$\ell_c$ is the de Broglie wavelength is that in the hard core
case all the energy is kinetic (the hard core just imposes a $\psi
=0$ boundary condition whenever the distance between two particles
is less than $a$). By the uncertainty principle, the kinetic
energy is proportional to an inverse length squared, namely
$\ell_c$. We then have the relation (since $\rho a ^3$ is small)
\begin{equation}\label{scales} a \ll \rho^{-1/3}\ll \ell_c \end{equation} 
which implies, physically, that
it is impossible to localize the particles relative to each other
(even though $\rho$ is small). Bosons in their ground state are
therefore `smeared out' over distances large compared to the mean
particle distance and their individuality is entirely lost. They
cannot be localized with respect to each other without changing
the kinetic energy enormously.

Fermions, on the other hand, prefer to sit in
`private rooms', i.e., $\ell_{c}$ is never bigger than $\rho^{-1/3}$
by a fixed factor.
In this respect the quantum nature of bosons is much more pronounced
than for fermions.

Since (\ref{basic}) is a basic result about the Bose gas it is clearly
important to derive it rigorously and in reasonable generality, in
particular for more general cases than hard spheres.  The question
immediately arises for which interaction potentials one may expect it
to be true. A notable fact is that it {\it not true for all} $v$ with
$a>0$, since there are two body potentials with positive scattering
length that allow many body bound states. (There are even such
potentials without two body bound states but with three body bound
states \cite{BA}.) For such potentials \eqref{basic} is clearly false.
Our proof, presented in the
sequel,  works for nonnegative $v$, but we conjecture that (\ref{basic})
holds if $a>0$ and $v$ has no $N$-body bound states for any $N$. The
lower bound is, of course, the hardest part, but the upper bound is not
altogether trivial either.

Before we start with the estimates a simple computation and some
heuristics may be helpful to make
(\ref{basic}) plausible and motivate the formal proofs.

With  $\psi_{0}$ the zero energy scattering solution,
partial integration, using \eqref{3dscatteq} and \eqref{3dscattlength}, 
gives, for $R\geq R_0$, 
\begin{equation}\label{partint}
\int_{|\x|\leq R}\{2\mu|\nabla \psi_{0}|^2+v|\psi_{0}|^2\}d\x=
8\pi\mu a\left(1-\frac aR\right)\to 8\pi\mu a\quad\mbox{\rm for $R\to\infty$}. 
\end{equation}
Moreover, for positive interaction potentials the scattering solution
minimizes
the quadratic form in (\ref{partint}) for each $R\geq R_0$ with
the boundary condition $\psi_0(|\x|=R)=(1-a/R)$. Hence the energy
$E_{0}(2,L)$ of two
particles in a large box, i.e., $L\gg
a$, is approximately $8\pi\mu a/L^3$. If the gas is sufficiently
dilute it is not unreasonable to expect that the energy is essentially
a sum of all such two particle contributions. Since there are
$N(N-1)/2$ pairs, we are thus lead to $E_{0}(N,L)\approx 4\pi\mu a
N(N-1)/L^3$, which gives (\ref{basic}) in the thermodynamic limit.

This simple heuristics is far from a rigorous proof, however,
especially for the lower bound. In fact, it is  rather remarkable that
the same asymptotic formula holds both for `soft' interaction
potentials, where perturbation theory can be expected to be a good
approximation, and potentials like hard spheres where this is not so.
In the former case the ground state is approximately the constant
function and the energy is {\it mostly potential}:
According to perturbation theory
$E_{0}(N,L)\approx  N(N-1)/(2 L^3)\int v(|\x|)d\x$. In particular it is {\it
independent of} $\mu$, i.e. of Planck's constant and mass. Since,
however, $\int v(|\x|)d\x$ is the first Born approximation to $8\pi\mu
a$ (note that $a$ depends on $\mu$!), this is not in conflict with
(\ref{basic}).
For `hard' potentials on the other hand, the ground state is {\it
highly correlated}, i.e., it is far from being a product of single
particle states. The energy is here {\it mostly kinetic}, because the
wave function is very small where the potential is large. These two
quite different regimes, the potential energy dominated one and the
kinetic energy dominated one, cannot be distinguished by the low
density asymptotics of the energy. Whether they behave
differently with respect to other phenomena, e.g., Bose-Einstein
condensation, is not known at present.

Bogolubov's analysis \cite{BO} presupposes the existence of Bose-Einstein
condensation. Nevertheless, it is correct (for the energy) for the
one-dimensional delta-function Bose gas \cite{LL}, despite the fact
that there is (presumably) no condensation in that case \cite{PiSt}. It turns
out that BE condensation is not really needed in order to understand
the energy.  As we shall see,  `global' condensation can be replaced
by a `local' condensation on boxes whose size is independent of
$L$. It is this crucial understanding that enables us to prove Theorem
1.1 without having to decide about BE condensation.

An important idea of Dyson was to transform the hard sphere
potential into a soft potential at the cost of sacrificing the
kinetic energy, i.e., effectively to move from one
regime to the other. We shall make use of this idea in our proof
of the lower bound below. But first we discuss the simpler upper
bound, which relies on other ideas from Dyson's beautiful paper \cite{dyson}.

\subsection{Upper Bound}\label{upsec}

The following generalization of Dyson's upper bound holds
\cite{LSY1999}, \cite{S1999}:
\begin{theorem}[\textbf{Upper bound}]\label{ub} Define $\rho_{1}=(N-1)/L^3$ and
$b=(4\pi\rho_{1}/3)^{-1/3}$. For nonnegative potentials $v$ and $b>a$
the ground state energy of (\ref{ham}) with periodic boundary conditions
satisfies
\begin{equation}\label{upperbound}
E_{0}(N,L)/N\leq 4\pi \mu \rho_{1}a\frac{1-\frac{a}
{b}+\left(\frac{a}
{b}\right)^2+\frac12\left(\frac{a}
{b}\right)^3}{\left(1-\frac{a}
{b}\right)^8}.
\end{equation}
For Dirichlet boundary conditions the estimate holds with ${\rm
(const.)}/L^2$ added to the right side.
Thus in the thermodynamic limit and for all boundary conditions
\begin{equation}
\frac{e_{0}(\rho)}{4\pi\mu\rho a}\leq\frac{1-Y^{1/3}+Y^{2/3}-\mfr1/2Y}
{(1-Y^{1/3})^8}.
\end{equation}
provided $Y=4\pi\rho a^3/3<1$.
\end{theorem}

\noindent {\it Remark.} The bound (\ref{upperbound}) holds for potentials
with infinite range, provided $b>a$. For potentials of finite range
$R_{0}$ it can be improved for $b>R_{0}$ to
\begin{equation}\label{upperbound2}
E_{0}(N,L)/N\leq 4\pi \mu \rho_{1}a\frac{1-\left(\frac{a}
{b}\right)^2+\frac12\left(\frac{a}
{b}\right)^3}{\left(1-\frac{a}
{b}\right)^4}.
\end{equation}

{\it Proof.}
We first remark that the expectation value of (\ref{ham}) with any
trial wave function gives an upper bound to the bosonic ground state
energy, even if the trial function is not symmetric under permutations
of the variables.  The reason is that an absolute ground state of the
elliptic differential operator (\ref{ham}) (i.e., a ground state
without symmetry requirement) is a nonnegative function which can be
be symmetrized without changing the energy because (\ref{ham}) is
symmetric under permutations.  In other words, the absolute ground
state energy is the same as the bosonic ground state energy.

Following \cite{dyson} we choose a trial function of
the following form
\begin{equation}\label{wave}
\Psi(\x_1,\dots,\x_N)=F_1(\x_{1}) \cdot F_2(\x_{1},\x_{2}) \cdots
F_N(\x_{1},\dots,\x_{N}).
\end{equation}
More specifically, $F_{1}\equiv 1$ and $F_{i}$ depends only on the
distance of $\x_{i}$ to its nearest neighbor among the the points
$\x_1,\dots ,\x_{i-1}$ (taking the periodic boundary into
account):
\begin{equation}\label{form}
F_i(\x_1,\dots,\x_i)=f(t_i), \quad t_i=\min\left(\xij,j=1,\dots,
i-1\right),
\end{equation}
with a function $f$ satisfying
\begin{equation}0\leq
f\leq 1, \quad f'\geq 0.
\end{equation}
The intuition behind the ansatz (\ref{wave}) is that the particles
are inserted into the system one at the time, taking into account
the particles previously inserted. While such a wave function
cannot reproduce all correlations present in the true ground
state, it turns out to capture the leading term in the energy for
dilute gases. The form (\ref{wave}) is  computationally easier to
handle than an ansatz of the type $\prod_{i<j}f(|\x_{i}-\x_{j}|)$,
which might appear more natural in view of the heuristic remarks
after Eq.\ \eqref{partint}.

The function $f$ is chosen to be
\begin{equation}\label{deff}
f(r)=\begin{cases}
f_0(r)/f_{0}(b)&\text{for $0\leq r\leq b$},\\
1&\textrm{for $r>b$},
\end{cases}
\end{equation}
with $f_{0}(r)=u_{0}(r)/r$ the zero energy scattering solution defined by 
\eqref{scatteq}. The estimates (\ref{upperbound}) and
(\ref{upperbound2}) are
obtained by somewhat lengthy computations similar as in
\cite{dyson}, but making use of
(\ref{partint}). For details we refer to \cite{LSY1999} and \cite{S1999}.

A test wave function with Dirichlet boundary condition may be obtained
by localizing the wave function (\ref{wave}) on the length scale $L$.
The energy cost per particle for this is ${\rm (const.)}/L^2$.
\hfill$\Box$

\subsection{Lower Bound}\label{subsect22}
It was explained previously in this section why the lower bound
for the bosonic ground state energy of (\ref{ham}) is not easy to  obtain.
The three different length scales \eqref{scales} for bosons will play a role in the
proof below.
\begin{itemize}
\item The scattering length $a$.
\item The mean particle distance $\rho^{-1/3}$.
\item The `uncertainty principle length' $\ell_{c}$, defined by
$\mu\ell_{c}^{-2}=e_{0}(\rho)$, i.e., $\ell_{c}\sim (\rho a)^{-1/2}$.
\end{itemize}

Our lower bound for $e_{0}(\rho)$ is as follows.
\begin{theorem}[\textbf{Lower bound in the thermodynamic limit}]\label{lbth}
For a  positive potential $v$ with finite range and $Y$ small enough
\begin{equation}\label{lowerbound}\frac{e_{0}(\rho)}{4\pi\mu\rho a}\geq
(1-C\,
Y^{1/17})
\end{equation}
with $C$ a constant. If $v$ does not have finite range, 
but decreases faster than  
$1/r^{3}$ (more precisely, $\int_R^\infty v(r)r^2 dr<\infty$ for some $R$) 
then an analogous
bound to (\ref{lowerbound})
holds, but with $CY^{-1/17}$ replaced by $o(1)$ as $Y\to 0$.
\end{theorem}
It should be noted right away that the error term $-C\, Y^{1/17}$ in
(\ref{lowerbound}) is of no fundamental significance and is
not believed to reflect the true state of affairs. Presumably, it
does not even have the right sign. We mention in passing that
$C$ can be taken to be
$8.9$ \cite{S1999}.

As mentioned at the beginning of this section after
Eq.\ \eqref{eq:thmlimit}, a lower bound on $E_{0}(N,L)$ for finite $N$
and $L$ is of importance for applications to inhomogeneous gases, and
in fact we derive (\ref{lowerbound}) from such a bound. We state it in
the following way:
\begin{theorem}[\textbf{Lower bound in a finite box}] \label{lbthm2}
    For a  positive potential $v$ with finite range there is
a $\delta>0$ such that the ground state energy of (\ref{ham}) with Neumann
boundary conditions satisfies
\begin{equation}\label{lowerbound2}E_{0}(N,L)/N\geq 4\pi\mu\rho
a \left(1-C\,
Y^{1/17}\right)
\end{equation}
for all $N$ and $L$ with $Y<\delta$ and $L/a>C'Y^{-6/17}$. Here $C$
and $C'$ are positive  constants, independent of $N$ and $L$. (Note that the
condition on $L/a$ requires in particular that $N$ must be large
enough, $N>\hbox{\rm (const.)}Y^{-1/17}$.)  As in Theorem \ref{lbth}
such a bound, but possibly with a different error term holds also for
potentials $v$ of infinite range that decrease sufficiently fast at
infinity.
\end{theorem}

The first step in the proof of Theorem \ref{lbthm2} is a generalization of
a lemma of Dyson, which allows us to replace $v$ by a `soft' potential,
at the cost of sacrificing kinetic energy and increasing the
effective range.

\begin{lemma}\label{dysonl} Let $v(r)\geq 0$ with finite range $R_{0}$. Let
$U(r)\geq 0$
be any function satisfying $\int U(r)r^2dr\leq 1$ and $U(r)=0$ for $r<R_{0}$.
Let
${\mathcal B}\subset \R^3$ be star shaped with respect to $0$ (e.g.\
convex with $0\in{\mathcal B}$). Then for all differentiable
functions $\psi$
\begin{equation}\label{dysonlemma}
    \int_{\mathcal B}\left[\mu|\nabla\psi|^2+\mfr1/2
v|\psi|^2\right]
\geq \mu a \int_{\mathcal B} U|\psi|^2.\end{equation}
\end{lemma}

\begin{proof}  
Actually, (\ref{dysonlemma}) holds with $\mu |\nabla \psi
(\x)|^2$
replaced by the (smaller) radial kinetic energy,
 $\mu |\partial \psi (\x)/ \partial r|^2$, and  it  suffices to
prove
the analog of (\ref{dysonlemma}) for the integral along each radial
line with fixed angular variables. Along such a line we write
$\psi(\x) = u(r)/r$ with $u(0)=0$. We consider first the special case
when when $U$ is a delta-function at some radius $R\geq
R_0$,
i.e., \begin{equation}\label{deltaU}U(r)=\frac{1}{
R^2}\delta(r-R).\end{equation}
For such $U$ the analog of (\ref{dysonlemma}) along the radial line is
\begin{equation}\label{radial}\int_{0}^{R_{1}}
    \{\mu[u'(r)-(u(r)/r)]^2+\mfr1/2v(r)|u(r)]^2\}dr\geq
    \begin{cases}
        0&\text{if $R_{1}<R$}\\
            \mu a|u(R)|^2/R^2&\text{if $R\leq R_{1}$}
\end{cases}
\end{equation}
where $R_{1}$ is the length of the radial line segment in ${\mathcal
B}$.
The case $R_{1}<R$ is trivial,
because $\mu|\partial \psi/\partial r|^2+\mfr1/2 v|\psi|^2\geq 0$.
(Note that positivity of $v$ is used here.) If $R\leq R_{1}$ we
consider the integral on the the left side of (\ref{radial}) from 0 to $R$
instead of $R_{1}$ and
minimize it under the boundary condition that $u(0)=0$
and $u(R)$ is a fixed constant. Since everything is homogeneous in $u$ we may
normalize this value to $u(R)=R-a$.
This minimization problem leads to the zero energy
scattering
equation (\ref{scatteq}). Since $v$ is positive, the
solution is a true minimum and not just a
stationary point.

Because $v(r)=0$ for $r>R_{0}$ the solution, $u_{0}$, satisfies $u_{0}(r)=r-a$
for $r>R_{0}$.  By partial integration,
\begin{equation}\int_{0}^{R}\{\mu[u'_{0}(r)-(u_{0}(r)/r)]^2+
    \mfr1/2v(r)|u_{0}(r)]^2\}dr=\mu a|R-a|/R\geq \mu
a|R-a|^2/R^2.
    \end
{equation}
But $|R-a|^2/R^2$ is precisely
the right side of (\ref{radial}) if $u$ satisfies the normalization condition.

This derivation of (\ref{dysonlemma}) for the special case (\ref{deltaU})
implies the
general case, because every $U$ can be written as a
superposition of  $\delta$-functions,
$U(r)=\int R^{-2}\delta(r-R)\,U(R)R^2 dR$, and $\int U(R)R^2 dR\leq 1$
by assumption.
\end{proof}

By dividing $\Lambda$ for given points $\x_{1},\dots,\x_{N}$ into Voronoi
cells  ${\mathcal B}_{i}$ that contain all points
closer to $\x_{i}$ than to $\x_{j}$ with $j\neq i$ (these
cells are star shaped w.r.t. $\x_{i}$, indeed convex), the
following corollary of Lemma \ref{dysonl} can be derived in the same
way as the corresponding  Eq.\ (28) in \cite{dyson}.

\begin{corollary}\label{2.6} For any $U$ as in Lemma \ref{dysonl}
\begin{equation}\label{corollary}H_{N}\geq \mu a W\end{equation}
with
\begin{equation}\label{W}W(\x_{1},\dots,\x_{N})=\sum_{i=1}^{N}U(t_{i}),
\end{equation}
where $t_{i}$ is the distance of $\x_{i}$ to its {\it nearest
neighbor} among the other points $\x_{j}$, $j=1,\dots, N$, i.e.,
\begin{equation}\label{2.29}t_{i}(\x_{1},\dots,\x_{N})=\min_{j,\,j\neq
i}|\x_{i}-\x_{j}|.\end{equation}
\end{corollary}
\noindent
(Note that $t_{i}$ has here a slightly different meaning than in
(\ref{form}), where it denoted the distance to the nearest neighbor
among the $\x_{j}$ with $j\leq i-1$.)

Dyson considers in \cite{dyson} a one parameter family of $U$'s that
is essentially the same as the following choice, which is convenient for the
present purpose:
\begin{equation}\label{softened}
U_{R}(r)=\begin{cases}3(R^3-R_{0}^3)^{-1}&\text{for
$R_{0}<r<R$ }\\
0&\text{otherwise.}
\end{cases}
\end{equation}
We denote the corresponding interaction (\ref{W}) by $W_R$. For the hard core
gas one obtains
\begin{equation}\label{infimum}E(N,L)\geq \sup_{R}\inf_{(\x_{1},\dots,\x_{N})}
\mu a
W_R(\x_{1},\dots,\x_{N})\end{equation}
where the infimum is over $(\x_{1},\dots,x_{N})\in\Lambda^{N}$ with
$|\x_{i}-\x_{j}|\geq R_{0}=a$,
because of the hard core. At fixed $R$ simple geometry gives
\begin{equation}\label{fixedR}\inf_{(\x_{1},\dots,\x_{N})}
W_R(\x_{1},\dots,\x_{N})\geq \left(\frac{A}{R^3}-\frac{B}{ \rho
R^6}\right)\end{equation}
with certain constants $A$ and $B$. An evaluation of these constants
gives Dyson's bound
\begin{equation}E(N,L)/N\geq \frac{1}{10\sqrt 2} 4\pi\mu \rho
a.\end{equation}

The main reason this method does not give a better bound is that
$R$ must be chosen quite big, namely of the order of the mean
particle distance $\rho^{-1/3}$, in order to guarantee that the
spheres of radius $R$ around the $N$ points overlap. Otherwise the
infimum of $W_R$ will be zero. But large $R$ means that $W_R$ is
small. It should also be noted that this method does not work for
potentials other than hard spheres: If $|\x_{i}-\x_{j}|$ is
allowed to be less than $R_{0}$, then the right side of
(\ref{infimum}) is zero because $U(r)=0$ for $r<R_{0}$.

For these reasons we take another route. We still use  Lemma
\ref{dysonl} to get into the soft potential regime, but we do {\it
not} sacrifice {\it all} the kinetic energy as in
(\ref{corollary}). Instead we write, for $\varepsilon>0$
\begin{equation}
    H_{N}=\varepsilon H_{N}+(1-\varepsilon)H_{N}\geq \varepsilon
T_{N}+(1-\varepsilon)H_{N}
\end{equation}
with $T_{N}=-\sum_{i}\Delta_{i}$ and use (\ref{corollary}) only for the
part $(1-\varepsilon)H_{N}$. This gives
\begin{equation}\label{halfway}H_{N}\geq \varepsilon T_{N}+(1-\varepsilon)\mu
a W_R.\end{equation} We consider the operator on the right side
from the viewpoint of first order perturbation theory, with
$\varepsilon T_{N}$ as the unperturbed  part, denoted $H_{0}$.

The ground state of $H_{0}$ in a box of side length $L$ is
$\Psi_{0}(\x_{1},\dots,\x_{N})\equiv L^{-3N/2}$ and we denote
expectation values in this state by $\langle\cdot\rangle_{0}$.
A  computation, cf.\ Eq.\ (21) in \cite{LY1998}, gives
\begin{eqnarray}\label{firstorder}4\pi\rho\left(1-\mfr1/N\right)&\geq&
\langle W_R\rangle_{0}/N  \nonumber   \\ &\geq& 4\pi\rho
\left(1-\mfr1/N\right)\left(1-\mfr{2R}/L\right)^3
\left(1+4\pi\rho(1-\mfr1/N)(R^3-R_{0}^3)/3)\right)^{-1}.\nonumber\\
\end{eqnarray}
The rationale behind the various factors is as follows: $(1-\mfr1/N)$
comes from the fact that the number of pairs is $N(N-1)/2$ and not
$N^2/2$, $(1-{2R}/L)^3$ takes into account the fact that the particles
do not interact beyond the boundary of $\Lambda$, and the last factor
measures the probability to find another particle within the
interaction range of the potential $U_R$ for a given particle.

The estimates (\ref{firstorder}) on the first order term look at first
sight quite promising, for if we let $L\to \infty$, $N\to \infty$ with
$\rho=N/L^3$ fixed, and subsequently take $R\to\infty$, then $\langle
W_R\rangle_{0}/N$ converges to $4\pi\rho$, which is just what is
desired.  But the first order result (\ref{firstorder}) is not a
rigorous bound on $E_0(N,L)$, we need {\it error estimates}, and these
will depend on $\varepsilon$, $R$ and $L$.

We now recall {\it Temple's inequality} \cite{TE} for the expectation
values of an operator $H=H_{0}+V$ in the ground state
$\langle\cdot\rangle_{0}$ of $H_{0}$. It is a simple
consequence of the operator inequality
\begin{equation}(H-E_{0})(H-E_{1})\geq 0\end{equation}
for the two lowest eigenvalues, $E_{0}<E_{1}$, of
$H$ and reads
\begin{equation}\label{temple}E_{0}\geq \langle H\rangle_{0}-\frac{\langle
H^2\rangle_{0}-\langle
H\rangle_{0}^2}{E_{1}-\langle H\rangle_{0}}\end{equation}
provided $E_{1}-\langle H\rangle_{0}>0$.
Furthermore, if $V\geq 0$ we may use $E_{1}\geq E_{1}^{(0)}$= second lowest
eigenvalue of $H_{0}$ and replace $E_{1}$ in (\ref{temple}) by $E_{1}^{(0)}$.

{}From (\ref{firstorder}) and (\ref{temple}) we get the estimate
\begin{equation}\label{estimate2}\frac{E_{0}(N,L)}{ N}\geq 4\pi \mu a\rho
\left(1-{\mathcal
E}(\rho,L,R,\varepsilon)\right)\end{equation}
with
\begin{eqnarray}\label{error}1-{\mathcal
E}(\rho,L,R,\varepsilon)&=&(1-\varepsilon)\left(1-\mfr1/{\rho
L^3}\right)\left(1-\mfr{2R}/L\right)^3
\left(1+\mfr{4\pi}/3\rho(1-\mfr1/N)(R^3-R_{0}^3))\right)^{-1}\nonumber\\
&\times&\left(1-\frac{\mu a\big(\langle
W_R^2\rangle_0-\langle W_R\rangle_0^2\big)}{\langle
W_R\rangle_0\big(E_{1}^{(0)}-\mu a\langle W_R\rangle_0\big)}\right).
\end{eqnarray}
To evaluate this further one may use the estimates (\ref{firstorder}) and the
bound
\begin{equation}\label{square}
\langle W_R^2\rangle_0\leq 3\frac N{R^3-R_0^3}\langle W_R\rangle_0
\end{equation}
which follows from $U_R^2=3({R^3-R_0^3})^{-1}U_R$ together with the
Cauchy-Schwarz
inequality. A glance at the form of the error term reveals, however, that it
is
{\it not} possible here to take the thermodynamic limit $L\to\infty$ with
$\rho$
fixed:
We have $E_{1}^{(0)}=\varepsilon\pi\mu/L^2$ (this is the kinetic energy of a
{\it single} particle in the first excited state in the box), and the factor
$E_{1}^{(0)}-\mu a\langle W_R\rangle_0$ in the denominator in (\ref{error})
is,
up to unimportant constants and lower order terms, $\sim (\varepsilon
L^{-2}-a\rho^2L^3)$. Hence the denominator eventually becomes negative and
Temple's inequality looses its validity if $L$ is large enough.

As a way out of this dilemma we divide the big box $\Lambda$ into cubic {\it
cells} of side length $\ell$ that is kept {\it fixed} as $L\to \infty$.  The
number of cells, $L^3/\ell^3$, on the other hand, increases with $L$.  The $N$
particles are distributed among these cells, and we use (\ref{error}), with
$L$
replaced by $\ell$, $N$ by the particle number, $n$, in a cell and $\rho$ by
$n/\ell^3$, to estimate the energy in each cell with {\it Neumann} conditions
on the boundary. 
For each distribution of the particles we add the
contributions from the cells, neglecting interactions across boundaries.
Since
$v\geq 0$ by assumption, this can only lower the energy.  Finally, we minimize
over all possible choices of the particle numbers for the various cells
adding up to $N$.  The energy obtained in this way is a lower bound to
$E_0(N,L)$,
because we are effectively allowing discontinuous test functions for the
quadratic form given by $H_N$.

In mathematical terms, the cell method leads to
\begin{equation}\label{sum}
E_0(N,L)/N\geq(\rho\ell^3)^{-1}\inf \sum_{n\geq 0}c_nE_0(n,\ell)
\end{equation}
where the infimum is over all choices of coefficients $c_n\geq 0$ (relative
number of cells containing exactly $n$ particles), satisfying the constraints
\begin{equation}\label{constraints}
\sum_{n\geq 0}c_n=1,\qquad \sum_{n\geq 0}c_n n=\rho\ell^3.
\end{equation}

The minimization problem for the distributions of the particles among the
cells would be easy if we knew that the ground state energy $E_0(n,\ell)$ (or
a
good
lower bound to it) were convex in $n$.  Then we could immediately conclude
that
it is best to have the particles as evenly distributed among the boxes as
possible, i.e., $c_n$ would be zero except for the $n$ equal to the
integer closest to
$\rho\ell^3$. This would give
\begin{equation}\label{estimate3}\frac{E_{0}(N,L)}{ N}\geq 4\pi \mu a\rho
\left(1-{\mathcal E}(\rho,\ell,R,\varepsilon)\right)\end{equation} i.e.,
replacement of $L$ in (\ref{estimate2}) by $\ell$, which is independent of
$L$.
The blow up of ${\mathcal E}$ for $L\to\infty$ would thus be avoided.

Since convexity of $E_0(n,\ell)$ is not known (except in the thermodynamic
limit)
we must resort to other means to show that $n=O(\rho\ell^3)$ in all
boxes. The rescue
comes from {\it superadditivity} of $E_{0}(n,\ell)$, i.e., the property
\begin{equation}\label{superadd}
 E_0(n+n',\ell)\geq E_0(n,\ell)+E_0(n',\ell)
\end{equation}
which follows immediately from $v\geq 0$ by dropping the interactions between
the $n$ particles and the $n'$ particles. The bound (\ref{superadd}) implies
in
particular that for any $n,p\in{\mathbb N}$ with $n\geq p$
\begin{equation}\label{superadd1}
E(n,\ell)\geq [n/p]\,E(p,\ell)\geq \frac n{2p}E(p,\ell)
\end{equation}
since the largest integer $[n/p]$ smaller than $n/p$ is in any case $\geq
n/(2p)$.

The way (\ref{superadd1}) is used is as follows:
Replacing $L$ by $\ell$, $N$ by $n$ and $\rho$ by $n/\ell^3$ in
(\ref{estimate2})  we have for fixed $R$ and $\varepsilon$
\begin{equation}\label{estimate4}
E_{0}(n,\ell)\geq\frac{ 4\pi \mu a}{\ell^3}n(n-1)K(n,\ell)
\end{equation}
with a certain function $K(n,\ell)$ determined by (\ref{error}).
We shall see that $K$ is monoton\-ously decreasing in $n$, so that
if $p\in{\mathbb N}$  and $n\leq p$ then
\begin{equation}\label{n<p}
E_{0}(n,\ell)\geq\frac{ 4\pi \mu a}{\ell^3}n(n-1)K(p,\ell).
\end{equation}
We now split the sum in (\ref{sum}) into two parts.
For $n<p$ we use (\ref{n<p}), and for $n\geq p$ we use (\ref{superadd1})
together with (\ref{n<p}) for $n=p$. The task is thus to minimize
\begin{equation}\label{task}
\sum_{n<p}c_n n(n-1)+\mfr1/2\sum_{n\geq p}c_nn(p-1)
\end{equation}
subject to the constraints ({\ref{constraints}).
Putting
\begin{equation}
k:=\rho\ell^3 \quad\text{and}\quad t:=\sum_{n<p}c_n n\leq k
\end{equation}
we have $\sum_{n\geq p}c_n n=k-t$, and since
$n(n-1)$ is convex in $n$ and vanishes for $n=0$, and $\sum_{n<p}c_n\leq 1$, the expression
(\ref{task})
is
\begin{equation}
\geq t(t-1)+\mfr1/2(k-t)(p-1).
\end{equation}
We have to minimize this for $1\leq t\leq k$. If $p\geq 4k$ the minimum is
taken
at $t=k$ and is equal to $k(k-1)$. Altogether we have thus shown that
\begin{equation}\label{estimate1}
\frac{E_{0}(N,L)}{ N}\geq 4\pi \mu a\rho\left(1-\frac1{\rho\ell^3} \right)
K(4\rho\ell^3,\ell).
\end{equation}

What remains is to take a closer look at $K(4\rho\ell^3,\ell)$,
which depends on the parameters $\varepsilon$ and $R$ besides
$\ell$, and choose the parameters in an optimal way.
{F}rom (\ref{error}) and (\ref{square}) we obtain
\begin{eqnarray}\label{Kformula}
K(n,\ell)&=&(1-\varepsilon) \left(1-\mfr{2R}/\ell\right)^3
\left(1+\mfr{4\pi}/3\rho(1-\mfr1/n)(R^3-R_{0}^3))\right)^{-1}
\nonumber
\\ &\times&\left(1-\frac3\pi
\frac{an}{(R^3-R_{0}^3)(\varepsilon\ell^{-2}-4a\ell^{-3}n(n-1))}\right).
\end{eqnarray}
The estimate (\ref{estimate4}) with this $K$ is valid as long as the
denominator in the last factor
in (\ref{Kformula}) is $\geq 0$, and in order to have a formula
for
all $n$ we can take 0 as a
trivial lower bound in other cases or when (\ref{estimate4}) is
negative. As required
for (\ref{n<p}), $K$ is monotonously decreasing in $n$. We now insert
$n=4\rho\ell^3$ and obtain
\begin{eqnarray}\label{Kformula2}
K(4\rho\ell^3,\ell)&\geq&(1-\varepsilon)\left(1-\mfr{2R}/\ell\right)^3
\left(1+({\rm const.})Y(\ell/a)^3 (R^3-R_{0}^3)/\ell^3\right)^{-1}
\nonumber
\\ &\times&\left(1-
\frac{\ell^3}{(R^3-R_{0}^3)}\frac{({\rm const.})Y}
{(\varepsilon(a/\ell)^{2}-({\rm const.})Y^2(\ell/a)^3)}\right)
\end{eqnarray}
with $Y=4\pi\rho a^3/3$ as before. Also, the factor
\begin{equation}
\left(1-\frac1{\rho\ell^3} \right)=(1-({\rm const.})Y^{-1}(a/\ell)^{3})
\end{equation}
in (\ref{estimate1})
(which is the ratio between
$n(n-1)$ and $n^2$) must not be be forgotten. We now make the ansatz
\begin{equation}\label{ans}
\varepsilon\sim Y^\alpha,\quad a/\ell\sim Y^{\beta},\quad
(R^3-R_{0}^3)/\ell^3\sim Y^{\gamma}
\end{equation}
with exponents $\alpha$, $\beta$ and $\gamma$ that we choose
in an optimal way. The conditions to be met are as follows:
\begin{itemize}
\item $\varepsilon(a/\ell)^{2}-({\rm const.})Y^2(\ell/a)^3>0$. This
holds for all small enough $Y$, provided
$\alpha+5\beta<2$ which follows from the conditions below.
\item $\alpha>0$ in order that $\varepsilon\to 0$ for $Y\to 0$.
\item $3\beta-1>0$ in order that  $Y^{-1}(a/\ell)^{3}\to 0$ for for $Y\to
0$.
\item $1-3\beta+\gamma>0$ in order that
$Y(\ell/a)^{3}(R^3-R_{0}^3)/\ell^3\to 0$ for for $Y\to 0$.
\item $1-\alpha-2\beta-\gamma>0$ to control the last factor in
(\ref{Kformula2}).
\end{itemize}
Taking
\begin{equation}\label{exponents}
\alpha=1/17,\quad \beta=6/17,\quad \gamma=3/17
\end{equation}
all these conditions are satisfied, and
\begin{equation}
\alpha= 3\beta-1=1-3\beta+\gamma=1-\alpha-2\beta-\gamma=1/17.
\end{equation}
It is also clear that
$2R/\ell\sim Y^{\gamma/3}=Y^{1/17}$, up to higher order terms.
This completes the proof of Theorems \ref{lbth} and \ref{lbthm2}, for
the case of potentials with finite range. By optimizing the
proportionality constants in (\ref{ans}) one can show that $C=8.9$ is
possible in Theorem \ref{lbth} \cite{S1999}. The extension to
potentials of infinite range but finite scattering length is obtained
by approximation by finite range potentials, controlling the change of
the scattering length as the cut-off is removed. See Appendix A in
\cite{LY2d} and Appendix B in \cite{LSY1999} for details. We remark
that a slower decrease of the potential than $1/r^3$ implies infinite
scattering length.
\hfill$\Box$\bigskip

The exponents (\ref{exponents}) mean in particular that
\begin{equation}a\ll R\ll \rho^{-1/3}\ll \ell \ll(\rho
a)^{-1/2},\end{equation}
whereas Dyson's method required $R\sim \rho^{-1/3}$ as already explained.
The condition $\rho^{-1/3}\ll \ell$ is required in order to have many
particles in each box and thus $n(n-1)\approx n^2$. The condition
$\ell \ll(\rho a)^{-1/2}$ is necessary for a spectral gap
gap $\gg e_{0}(\rho)$ in Temple's inequality. It is also clear that
this choice of $\ell$  would lead to a far too big
energy and no bound for $ e_{0}(\rho)$ if we had chosen Dirichlet instead of
Neumann boundary
conditions for the cells. But with the latter the method works!

\bigskip

\section{The Dilute Bose Gas in 2D} \label{sect2d}

In contrast to the three-dimensional theory, the two-dimensional Bose
gas began to receive attention only relatively late.  The first
derivation of the correct asymptotic formula was, to our knowledge,
done by Schick \cite{schick} for a gas of hard discs. He found
\begin{equation}
    e(\rho)  \approx 4\pi \mu \rho |\ln(\rho a^2) |^{-1}.
\label{2den}
\end{equation}
This was accomplished by an infinite summation of `perturbation series'
diagrams. Subsequently, a corrected modification of \cite{schick} was
given in \cite{hines}. Positive temperature extensions were given in
\cite{popov} and in \cite{fishho}. All this work involved an analysis in
momentum space, with the exception
of a method due to one of us that works directly in configuration space
\cite{Lieb63}.  Ovchinnikov \cite{Ovch} derived \eqref{2den} by using,
basically, the method in \cite{Lieb63}. These derivations require
several unproven assumptions and are not rigorous. 

In two dimensions the scattering length $a$ is defined using the zero
energy scattering equation (\ref{3dscatteq}) but instead of
$\psi(r)\approx 1-a/r$ we now impose the asymptotic condition
$\psi(r)\approx \ln(r/a)$.
This is explained  in the appendix to \cite{LY2d}.

Note that in two dimensions the ground state energy  
could not possibly be $e_0(\rho)  \approx 4\pi \mu
\rho a$ as in three dimensions because that would be dimensionally wrong. 
Since $e_0(\rho) $ should essentially be proportional to $\rho$,
there is apparently no room for an $a$ dependence --- which is
ridiculous! It turns out that this dependence comes about in the
$\ln(\rho a^2)$ factor.

One of the intriguing facts about \eqref{2den} is that the energy for $N$
particles is {\it not equal} to $N(N-1)/2$  times the energy for two
particles in the low density limit --- as is the case in
three dimensions.  The latter quantity,  $E_0(2,L)$, is, asymptotically
for large $L$, equal to $8\pi \mu L^{-2} \left[ \ln(L^2/a^2)
\right]^{-1}$.  (This is seen in an analogous way as \eqref{partint}. The 
three-dimensional boundary condition $\psi_0(|\x|=R)=1-a/R$ is
replaced by $\psi_0(|\x|=R)=\ln (R/a)$ and moreover it has to be taken
into account that with this normalization $\|\psi_0\|^2={\rm
(volume)}(\ln (R/a))^2$ (to leading order), instead of just the volume
in the three-dimensional case.)  Thus, if the $N(N-1)/2$ rule were to
apply, \eqref{2den} would have to be replaced by the much smaller
quantity $4\pi \mu
\rho\left[ \ln(L^2/a^2) \right]^{-1}$. In other words, $L$, which tends
to $\infty$ in the thermodynamic limit, has to be replaced by the mean
particle separation, $\rho^{-1/2}$ in the logarithmic factor. Various
poetic formulations of this curious fact have been given, but the fact
remains that the non-linearity is something that does not  occur in more
than two dimensions and its precise nature is hardly obvious,
physically. This anomaly is the main reason that the two-dimensional case is
not a trivial extension of the three-dimensional one.

Eq.\ \eqref{2den} was proved in \cite{LY2d} for nonnegative, finite
range two-body potentials by finding upper and lower bounds of the
correct form, using similar ideas as in the previous section for the
three-dimensional case. We discuss below the modifications that have to
be made in the present two-dimensional case.  The restriction to
finite range can be relaxed as in three dimensions, but the
restriction to nonnegative $v$ cannot be removed in the current state
of our methodology.  The upper bounds will have relative remainder
terms O($|\ln(\rho a^2)|^{-1}$) while the lower bound will have
remainder O($|\ln(\rho a^2)|^{-1/5}$).  It is claimed in \cite{hines}
that the relative error for a hard core gas is negative and O$(\ln
|\ln(\rho a^2)||\ln(\rho a^2)|^{-1})$, which is consistent with our
bounds.

The upper bound is derived in complete analogy with the three
dimensional case. The function $f_0$ in the variational ansatz 
\eqref{deff} is in
two dimensions also the zero energy scattering solution --- but for 
2D, of course. The result is 
\begin{equation}\label{upperbound3}
E_{0}(N,L)/N\leq \frac{2\pi\mu\rho}{\ln(b/a)-\pi\rho b^{2}}\left(1+{\rm 
O}([\ln(b/a)]^{{-1}})\right).
\end{equation}
The minimum over $b$ of the leading term is obtained for 
$b=(2\pi\rho)^{{-1/2}}$. Inserting this in \eqref{upperbound3} we thus 
obtain
\begin{equation}\label{upperbound1}
E_{0}(N,L)/N\leq \frac{4\pi\mu\rho}{|\ln(\rho a^{2})|}\left(1+{\rm 
O}(|\ln(\rho a^{2})|^{{-1}}\right).
\end{equation}

To prove the lower bound the essential new step is to modify Dyson's lemma
for 2D. The 2D version of
Lemma \ref{dysonl} is:

\begin{lemma}\label{dyson2d} 
Let $v(r)\geq0$ and $v(r)=0$ for $r>R_0$.
 Let $U(r)\geq 0$ be any function satisfying
\begin{equation}\label{1dyson}
 \int_0^\infty U(r)\ln(r/a)rdr \leq 1~~~~~{\rm and}~~~~~ U(r)=0 ~~~{\rm
for}~
 r<R_0.
\end{equation}
Let ${\mathcal B}\subset \R^2$ be star-shaped  with respect
to $0$ (e.g.\ convex
with $0\in{\mathcal B}$).
Then, for all functions $\psi$ in the Sobolev space $H^1(\mathcal{B})$,
\begin{equation}
\int_{\mathcal B} \left(\mu|\nabla \psi (\x)|^2 + \half v(|\x|)
|\psi (\x)|^2\right)~d\x  
\geq  \mu  \int_{\mathcal B} U(|\x|)
|\psi (\x)|^2 ~d\x.
\label{dysonlem2d}
\end{equation}
\end{lemma}

\begin{proof}
In polar coordinates, $r,\theta$, one has 
$|\nabla \psi|^2 \geq |\partial \psi /\partial r|^2$. Therefore, it 
suffices to 
prove that for each angle $\theta \in [0,2\pi)$, and with
$\psi (r,\theta)$ denoted simply by $f(r)$,
\begin{equation} \label{radial2}
\int_0^{R(\theta)}\left( \mu |\partial f(r) /\partial r|^2 + 
\half v(r)|f(r)|^2 \right)rdr \geq  
 \mu  \int_0^{R(\theta)}  U(r)|f(r)|^2 ~rdr,
\end{equation}  
where $R(\theta)$ denotes the distance of the origin to the boundary
of $\mathcal{B}$ along the ray $\theta$. 

If $R(\theta) \leq R_0$ then \eqref{radial2} is trivial because the 
right side is zero while the left side is evidently nonnegative. (Here,
$v\geq0$ is used.)

If $R(\theta) > R_0$ for some given
value of $\theta$, consider the disc $\mathcal{D}(\theta)=
\{\x\in \mathbb{R}^2 \ :\ 0\leq |\x|\leq R(\theta) \}$ centered at the
origin in $\mathbb{R}^2$ and of radius $R(\theta) $.
Our function $f$ defines a spherically
symmetric function, $\x\mapsto f(\vert \x\vert)$ on
$\mathcal{D}(\theta)$, and \eqref{radial2} is 
equivalent to 
\begin{equation}\label{disc}
\int_{{\mathcal D}(\theta)} \left(\mu|\nabla f (|\x|)|^2 + \frac{1}{2}v(|\x|)
|f(|\x|)|^2\right)d\x\geq 
\mu\int_{{\mathcal D}(\theta)} U(|\x|)|f(|\x|)|^2 d\x.
\end{equation}

Now choose some $R\in (R_0,\ R(\theta))$ and note that the left side of
\eqref{disc} is not smaller than the same quantity with 
${\mathcal D}(\theta)$ replaced by the smaller disc ${\mathcal D}_R=
\{\x\in \mathbb{R}^2 \ :\ 0\leq |\x|\leq R \}$. (Again, $v\geq 0$ is used.)
We now minimize this integral over ${\mathcal D}_R$, fixing $f(R)$. 
This minimization problem leads to the zero energy scattering equation. 
Plugging in the solution and integrating by parts leads to
\begin{equation} \label{pointwise}
2\pi \int_0^{R(\theta)}\left( \mu |\partial f(r) /\partial r|^2 + 
\frac{1}{2}v(r)|f(r)|^2 \right)rdr \geq    \frac{2\pi \mu}{\ln (R/a)}|f(R)|^2 .
\end{equation}
The proof is completed
by multiplying  both sides of \eqref{pointwise} by $U(R)R\ln(R/a)$ and 
integrating with respect to $R$ from $R_0$ to $R(\theta)$.
\end{proof}

As in Corollary \ref{2.6}, Lemma \ref{dyson2d} can be used to bound the 
many body Hamiltonian $H_N$ from below, as follows:
\begin{corollary}
For any $U$ as in Lemma \ref{dyson2d} and any $0<\varepsilon<1$
\begin{equation}\label{epsilonbd}
H_N \geq \varepsilon T_N +(1-\varepsilon)\mu W
\end{equation}
with $T_N=-\mu\sum_{i=1}^{N}\Delta_{i}$ and 
\begin{equation}
\label{W2}W(\x_{1},\dots,\x_{N})=\sum_{i=1}^{N}U\left(\min_{j,\,j\neq
i}|\x_{i}-\x_{j}|.\right).
\end{equation}
\end{corollary} 

For $U$ we choose the following functions, parameterized by $R>R_{0}$:
\begin{equation}U_{R}(r)=\begin{cases}\nu(R)^{-1}&\text{for
$R_{0}<r<R$ }\\
0&\text{otherwise}
\end{cases}
\end{equation}
with $\nu(R)$ chosen so that
\begin{equation}
\int _{R_{0}}^{R}U_{R}(r)\ln(r/a)r\,dr=1
\end{equation}
for all $R>R_{0},$
i.e.,
\begin{equation}\label{nu}
\nu(R)=\int_{R_{0}}^{R}\ln(r/a)r\,dr=\mfr1/4 \left\{R^{2}
\left(\ln(R^{2}/a^{2})-1\right)-R_{0}^{2}
\left(\ln(R_{0}^{2}/a^{2})-1\right)\right\}.
\end{equation}
The nearest neighbor 
interaction \eqref{W2} corresponding to $U_{R}$ will be denoted $W_{R}$.

As in Subsection \ref{subsect22} we shall need estimates  on the expectation
value, $\langle W_R\rangle_{0}$,  of $W_{R}$ in the ground state of
$\varepsilon T_N$ of \eqref{epsilonbd} with
Neumann boundary conditions. This is just the average value of $W_{R}$
in a hypercube in $\R^{2N}$.  Besides the
normalization factor $\nu(R)$, the computation involves 
the volume (area) of the support of
$U_{R}$, which is
 \begin{equation} A(R)=\pi(R^{2}-R_{0}^{2}).
\end{equation}

In contrast to the three-dimensional situation the normalization factor
$\nu(R)$ is not just a constant ($R$ independent) multiple of $A(R)$;
 the factor $\ln(r/a)$ in \eqref{1dyson} accounts for the more
complicated expressions in the two-dimensional case.   Taking into
account that $U_{R}$ is proportional to the characteristic function of 
a disc of radius $R$ with a hole of radius $R_{0}$, the following 
inequalities for $n$ particles in a box of side length 
$\ell$ are obtained by the same geometric reasoning as lead to 
\eqref{firstorder}, cf.\  
\cite{LY1998}:
\begin{eqnarray}\label{2dfirstorder}
\langle 
W_R\rangle_{0}&\geq&\frac {n}{\nu(R)}
\left(1-\mfr {2R}/{\ell}\right)^2\left[1-(1-Q)^{(n-1)}\right]\\
\langle 
W_R\rangle_{0}&\leq&
\frac 
{n}{\nu(R)}\left[1-(1-Q)^{(n-1)}\right]
\end{eqnarray}
with 
\begin{equation}
Q=A(R)/\ell^{2}
\end{equation}
  being   the relative volume occupied by the
support of the potential $U_{R}$.
Since $U_{R}^{2}=\nu(R)^{-1}U_{R}$ we also have
\begin{equation}
\langle 
W_R^{2}\rangle_{0}\leq \frac n{\nu(R)}\langle 
W_R\rangle_{0}.
\end{equation}

As in \cite{LY1998} we estimate $[1-(1-Q)^{(n-1)}]$ by
\begin{equation}
(n-1)Q\geq \left[1-(1-Q)^{(n-1)}\right]\geq \frac{(n-1)Q}{1+(n-1)Q}
\end{equation}
This gives
\begin{eqnarray}\label{firstorder2}
\langle 
W_R\rangle_{0}&\geq&\frac {n(n-1)}{\nu(R)}
\cdot\frac{Q}{1+(n-1)Q},\\
\langle 
W_R\rangle_{0}&\leq&
\frac {n(n-1)}{\nu(R)}
\cdot Q\ .
\end{eqnarray}

{}From  Temple's inequality \cite{TE} we 
obtain like in \eqref{temple} the estimate
\begin{equation}\label{temple2d}E_{0}(n,\ell)
\geq (1-\varepsilon)\langle 
W_R\rangle_{0}\left(1-\frac{\mu \big(\langle
W_R^2\rangle_0-\langle W_R\rangle_0^2\big)}{\langle
W_R\rangle_0\big(E_{1}^{(0)}-\mu \langle W_R\rangle_0\big)}\right)
\end{equation}
where
\begin{equation}
E_{1}^{(0)}=\frac{\varepsilon\mu}{\ell^{2}}
\end{equation}
is the energy of the lowest excited state of $\varepsilon T_n$. 
This estimate is valid for $E_{1}^{(0)}/\mu > \langle 
W_R\rangle_0$, i.e., it is important that $\ell$ is not too big. 

Putting \eqref{firstorder2}--\eqref{temple2d} together we obtain 
the estimate
\begin{equation}\label{alltogether}E_{0}(n,\ell)\geq 
\frac{n(n-1)}{\ell^{2}}\,\frac {A(R)}{\nu(R)}\,
K(n)
\end{equation}
with
\begin{equation}\label{k}
K(n)=(1-\varepsilon)\cdot 
\frac{(1-\mfr{2R}/{\ell})^{2}}{1+(n-1)Q}\cdot
\left(1-\frac n{(\varepsilon\,\nu(R)/\ell^{2})-n(n-1)\,Q}\right)
\end{equation}
Note that $Q$ depends on $\ell$ and $R$, and $K$ depends on 
$\ell$, $R$ and $\varepsilon$ besides $n$.  We have here dropped the 
term  $\langle W_R\rangle_0^2$ in the numerator in \eqref{temple2d},  
which is  appropriate  for the purpose of a lower bound. 

We note that $K$ is monotonically decreasing in $n$, so for a given 
$n$ we may replace $K(n)$ by $K(p)$ provided $p\geq n$.  As 
explained in the previous section, \eqref{superadd}--\eqref{estimate1}, 
convexity of $n\mapsto n(n-1)$ together with 
superadditivity of $E_{0}(n,\ell)$ in $n$ 
leads, for $p=4\rho\ell^{2}$,  to an estimate for the energy of $N$ 
particles in the 
large box when the  side length $L$ is  an integer multiple of $\ell$: 
\begin{equation}\label{alltogether2}E_{0}(N,L)/N\geq \frac 
{\rho A(R)}{\nu(R)}\left (1-\frac 1{\rho\ell^{2}}\right) 
K(4\rho\ell^{2})
\end{equation}
with $\rho=N/L^2$.

Let us now look at the conditions on the parameters $\varepsilon$, $R$ 
and $\ell$ that have to be met in order to 
obtain a lower bound with the same leading term as the upper bound 
\eqref{upperbound1}.

{}From \eqref{nu} we have 
\begin{equation}\label{alltogether3}
\frac{A(R)}{\nu(R)}=\frac{4\pi}
{\left(\ln(R^{2}/a^{2})-1\right)}\left(1-{\rm 
O}((R_{0}^{2}/R^{2})\ln(R/R_{0})\right)
\end{equation}
We thus see that as long as $a<R<\rho^{-1/2}$ the logarithmic factor
in the denominator in \eqref{alltogether3} has the right form for a lower 
bound. Moreover, for 
Temple's inequality the denominator in the third factor in \eqref{k} 
must be positive. With $n=4\rho\ell^2$ and 
        $\nu(R)\geq {\rm(const.)} R^2\ln(R^2/a^2)\ {\rm for}\ R\gg R_{0}$,
this condition amounts to
\begin{equation}\label{templecond}
{\rm (const.)}\varepsilon \ln(R^2/a^2) /\ell^{2}>\rho^{2}\ell^{4}.
\end{equation}
The relative error terms in \eqref{alltogether2} that have to be $\ll 1$ 
are
\begin{equation}\label{errors}
        \varepsilon,\quad \frac{1}{\rho\ell^{2}},\quad 
\frac{R}{\ell},\quad\rho R^2,\quad
\frac{\rho\ell^4}{\varepsilon R^2\ln(R^2/a^2)}.
\end{equation}
We now choose
\begin{equation}
\varepsilon\sim|\ln(\rho a^2)|^{-1/5},
\quad \ell\sim \rho^{-1/2}|\ln(\rho a^2)|^{1/10},
\quad R\sim \rho^{-1/2}|\ln(\rho a^2)|^{-1/10}
\end{equation}

Condition \eqref{templecond} is satisfied since the left side is $>{\rm
(const.)}|\ln(\rho a^2)|^{3/5}$ and the right side is $\sim |\ln(\rho
a^2)|^{2/5}$. The first three error terms in \eqref{errors} are all of
the same order, $|\ln(\rho a^2)|^{-1/5}$, the last is $\sim
|\ln(\rho a^2)|^{-1/5}(\ln|\ln(\rho a^2)|)^{-1}$. With these 
choices, \eqref{alltogether2} thus leads to the following:

\begin{theorem}[{\bf Lower bound}]
For all $N$ and $L$ large enough such that $L>{\rm (const.)}
\rho^{-1/2}|\ln(\rho a^2)|^{1/10}$ and 
$N>{\rm (const.)}|\ln(\rho a^2)|^{1/5}$ with $\rho=N/L^2$, the 
ground state energy with Neumann boundary condition satisfies
\begin{equation} \label{lower}
E_{0}(N,L)/N\geq \frac{4\pi\mu\rho}{|\ln(\rho a^2)|}
\left(1-{\rm O}(|\ln(\rho a^2)|^{-1/5})\right). 
\end{equation}  
\end{theorem}

In combination with the upper bound \eqref{upperbound1} this also proves
\begin{theorem}[{\bf Energy at low density in the thermodynamic limit}]
\begin{equation}
\lim_{\rho a^2\to 0}\frac{e_0(\rho)}{4\pi\mu\rho|\ln(\rho a^2)|^{-1}}=1
\end{equation}
where $e_0(\rho)=\lim_{N\to\infty} E_0(N,\rho^{-1/2}N^{1/2})/N$.
This holds irrespective of boundary conditions.
\label{limitthm}
\end{theorem}

As in the three-dimensional case, Theorem \ref{limitthm} is also valid
for an infinite range potential $v$ provided that $v\geq 0$ and for
some $R$ we have $\int_{R}^{\infty} v(r)r\ dr <\infty$, which
guarantees a finite scattering length.

\bigskip

\section{Bose-Einstein Condensation}\label{sectbe}

Let us now comment on the notion of Bose-Einstein condensation
(BEC). Given the normalized ground state wave function
$\Psi_{0}(\x_{1},\dots,\x_{N})$ we can form the one-body density
matrix which is an operator on $L^2(\R^d)$    ($d=2$ or $3$) given
by the kernel
\begin{equation}\label{defgamma}
 \gamma(\x,\x')=N\int \Psi_0(\x,\X)
\Psi_0(\x',\X) d\X \ ,
\end{equation}
 where we introduced the short hand
notation
\begin{equation}\label{defX}
\X=(\x_2,\dots,\x_N)\qquad{\rm and}\quad  d\X=\prod\limits_{j=
2}^N d\x_j.
\end{equation}
Then $\int \gamma(\x,\, \x) d\x =\Tr[\gamma] = N$. BEC in the ground state 
is the
assertion that this operator has an eigenvalue of order $N$ in 
the thermodynamic limit. Since
$\gamma$ is  a positive kernel and, hopefully, translation
invariant in the thermodynamic limit, the eigenfunction belonging
to  the largest eigenvalue must be the constant function
$L^{-d/2}$. Therefore, another way to say that there is BEC in the ground 
state is that
\begin{equation}\label{defbec}
 \frac 1{L^d} \int\int \gamma(\x,\, \y) d\x
d\y = \textrm{O}(N)\ 
\end{equation}
as $N\to \infty$, $L\to \infty$ with $N/L^d$ fixed. 
Unfortunately, this is something that is frequently invoked but never
proved for many body Hamiltonians with genuine interactions 
--- except for one special case:  hard core bosons on a
lattice at half-filling (i.e.,
$N=$ half the number of lattice sites). The proof is in \cite{KLS}.

The problem remains open after more than 75 years since the first 
investigations on 
the Bose gas \cite{Bose,Einstein}. It is also not at all clear
that BEC is essential for superfluidity, as frequently claimed. Our
construction in Section \ref{sect3d} shows that (in 3D) BEC exists on
a length scale of order $\rho^{-1/3} Y^{ -1/17}$ which, unfortunately,
is not a `thermodynamic' length like $\textrm{volume}^{1/3}$. The same
remark applies to the 2D case of Section 3, where BEC is proved over a
length scale $\rho^{-1/10}|\ln(\rho a^2)|^{1/10}$.

In a certain {\it dilute} limit, however, one can prove
(\ref{defbec}), as has been recently shown in 
\cite{LS02}. In this limit the interaction potential $v$ 
is varied with $N$ so that the ratio $a/L$ of the scattering length to
the box length is of order $1/N$. In \cite{LS02} the case of a Bose
gas confined in an external trap potential was considered (see
Sections \ref{sectgp} and \ref{becsect}), but the analysis for a
homogeneous gas is even simpler and implies the following result.  For
simplicity, we shall treat only the 3D case.

\begin{theorem}[\textbf{BEC in a dilute limit}]\label{hombecthm}
Assume that, as $N\to\infty$, $\rho=N/L^3$ and $\g=Na/L$ stay
fixed, and impose either periodic or Neumann boundary conditions
for $H$. Then
\begin{equation}
\lim_{N\to\infty} \frac 1N \frac 1{L^3} \int\int \gamma(\x,\, \y)
d\x d\y = 1\ .
\end{equation}
\end{theorem}

The reason why the limit $N\to\infty$ with $Na/L$ fixed is
particularly interesting will become clear when we study systems
confined in a trap potential in the next section. Note that the
limit we consider is really a limit of a {\it dilute} gas, since
$$a^3\rho = \left(\frac{Na}L\right)^3 \frac 1{N^2}=O(N^{-2})$$ as
$N\to\infty$. Since the ground state energy is of the order
$a\rho\sim L^{-2}$ is this limit, it is also clear why we do not
deal with Dirichlet boundary conditions: there would be an
additional contribution to the energy of the same order, and the
system would not be homogeneous any more. Dirichlet boundary
conditions can, however, be treated with the methods of Section
\ref{becsect}.

At this point we should say what we mean exactly by changing $a$
with $N$. We do this by scaling, i.e., we write
\begin{equation}\label{v1}
v(|\x|)=\frac 1{a^2} v_1(|\x|/a)
\end{equation}
for some $v_1$ having scattering length $1$, and vary $a$ while
keeping $v_1$ fixed. It is easily checked that the $v$ so defined has
scattering length $a$. It is important to note that, in
 the limit considered,  $a$ tends to zero (as
$N^{-2/3}$), and $v$ becomes a {\it hard} potential of {\it short}
range. This is the {\it opposite} of the usual mean field limit where
the strength of the potential goes to zero while its range tends to
infinity.

\begin{proof}[Proof of Theorem \ref{hombecthm}]
Let $\g=Na/L$ and $\rho=N/L^3$ be fixed as $N\to\infty$. Since
$a^3\rho\to 0$, Theorems \ref{ub} and 
\ref{lbthm2} imply 
\begin{eqnarray}\nonumber
\lim_{N\to\infty} \left(\frac N\rho\right)^{2/3} &&\!\!\!\!\!\!\!\!\!\!
\int d\X d\x_1
\Biggl(  \mu|\nabla_{\x_1} \Psi_0(\x_1,\X)|^2
\\ && + \half\sum_{j=2}^N
v(|\x_1-\x_j|)|\Psi_0(\x_1,\X)|^2\Biggl) \label{refor} = 4\pi\mu\g,
\end{eqnarray}
where we again used the short hand notation (\ref{defX}). 
The symmetry of $\Psi_0$ and the boundary conditions have also been used. 
Even
more is true, namely that
\begin{equation}\label{45}
\lim_{N\to\infty} \left(\frac N\rho\right)^{2/3} \int d\X d\x_1
\mu|\nabla_{\x_1} \Psi_0(\x_1,\X)|^2 = 4\pi\mu\g s
\end{equation}
and
\begin{equation}
\lim_{N\to\infty} \left(\frac N\rho\right)^{2/3} \int d\X d\x_1
\half\sum_{j=2}^N v(|\x_1-\x_j|)|\Psi_0(\x_1,\X)|^2 = 4\pi\mu\g
(1-s)
\end{equation}
for some $0<s\leq 1$. The parameter $s$ is given by
$s=\int|\nabla\psi_0|^2/(4\pi a)$, where $\psi_0$ denotes the
solution to the scattering equation for $v$ (under the boundary
condition $\lim_{|\x|\to\infty}\psi_0(\x)=1$; see Eq.
(\ref{3dscatteq})). This is a simple consequence of (\ref{refor}) by
variation with respect to the different components of the energy, as
was also noted in \cite{CS01a}. More precisely, the ground state
energy is a concave function of the mass parameter $\mu$, so it is
legitimate to differentiate both sides of \eqref{refor} with respect
to $\mu$. In doing so, it has to be noted that $g$ depends on $\mu$
through the scattering length. Using \eqref{partint} one sees that
\begin{equation}\label{dmua}
\frac{d(\mu a)}{d\mu}=\frac1 {4\pi}\int |\nabla\psi_0|^2d\x
\end{equation}
by the Feynman-Hellmann principle, since $\psi_0$ minimizes the left
side of \eqref{partint}.

We now concentrate on the term (\ref{45}) and show that to leading
order all the energy  is located in small balls surrounding each
particle. These balls can be taken to have radius roughly
$N^{-4/51}$ compared to the mean particle distance $\rho^{-1/3}$.
More precisely, we will show that
\begin{equation}\label{47}
\lim_{N\to\infty} N^{2/3} \int d\X \int_{\Omega_\X} d\x_1
|\nabla_{\x_1} \Psi_0(\x_1,\X)|^2 = 0\ ,
\end{equation}
where, for fixed $\X$, $\Omega_\X$ is given by
\begin{equation}
\Omega_\X=\left\{\x\in \Lambda \left| \, \min_{k\geq
2}|\x-\x_k|\geq N^{-4/51}\right\}\right. .
\end{equation}

To see this, we shall show that (\ref{refor}) still holds true if
the integrals in the first term are restricted to the complement
of $\Omega_\X$, denoted by $\Omega_\X^c$. The proof of this is
actually just a detailed examination of the lower bounds to the
ground state energy derived in Subsection \ref{subsect22}. What we
have to show is
\begin{equation}\label{qfff}
\sum_{i=1}^{N} \int_{\Omega_i^c} \mu |\nabla_i
\Psi_0|^2 d\x_1 d\X +\sum_{ i<j} \int
v(|\x_i-\x_j|)|\Psi_0|^2d\x_1 d\X  \geq 4\pi\mu a \rho N (1-o(1))
\end{equation}
as $N\to\infty$, where $\Omega_i^c$ denotes the set
$$\Omega_i^c=\{(\x_1,\X)\in\Lambda^{N}| \, \min_{k\neq
i}|\x_i-\x_k|\leq N^{-4/51}\}\ .$$

While  (\ref{qfff}) is not true for all conceivable $\Psi$'s
satisfying the normalization condition $\|\Psi\|_2=1$, it {\it is}
true for $\Psi_0$. Namely, we claim that
\begin{eqnarray}\nonumber &&\sum_{i=1}^{N} \left(\int_{\Omega_i^c} \mu |\nabla_i
\Psi|^2 d\x_1 d\X+\eps \int |\nabla_i \Psi|^2 d\x_1 d\X\right)
\\ \label{qffff}  &&+\sum_{ i<j} \int v(|\x_i-\x_j|)|\Psi|^2d\x_1 d\X
 \geq 4\pi\mu a \rho N (1-o(1))
\end{eqnarray}
for any $\Psi$, as long as $\eps\geq O(N^{-2/17})$ as $N\to\infty$. 
Since (\ref{45}) implies that $\Psi_0$ has
total kinetic energy of order $O(a\rho N)=O(N^{1/3})$,
(\ref{qfff}) follows from (\ref{qffff}).

It remains to prove (\ref{qffff}), but this has essentially
already been done in Subsection \ref{subsect22}! Namely, the usage
of the Dyson Lemma \ref{dysonl} requires only the kinetic energy
inside balls of radius $R=O(N^{-4/51})$ around each particle. In fact, 
by Eqs.\eqref{ans} and \eqref{exponents}, $R\sim aY^{-5/17}$ with $Y\sim a^3\rho\sim N^{-2}$ and $a\sim N^{-2/3}$ 
(the latter because $g=Na/L$ is fixed and $L\sim N^{1/3}$).  
The
second term in (\ref{qffff}), the total kinetic energy multiplied
by $\eps$, together with the `softened' potential (\ref{softened}),
then gives the desired lower bound to the energy (see
(\ref{halfway})--(\ref{estimate2})), as long as $\eps\geq
O(N^{-2/17})$. This proves (\ref{qffff}) 
with an error term of the order $O(N^{-2/17})$, and therefore also 
\eqref{qfff} with a relative error $O(N^{-2/17})$.

Eq. (\ref{47}) means that $\nabla_\x
\Psi_0(\x,\X)$ is almost zero (in an $L^2$ sense) 
outside of the small balls $\Omega_\X^c$. To conclude BEC we need to
show that as a function of $\x$, $\Psi_0(\x,\X)$ is essentially
constant in $\Omega_\X$. Although $\Omega_\X$ has a big volume, it can
be a weird, and even disconnected, set, so this conclusion is not yet
possible. However, exploiting the knowledge that the {\it total}
kinetic energy of $\Psi_0$ (including the balls) is not huge (see
(\ref{45})), we an show the desired constancy of $\Psi_0$. What we
need for this purpose is the following special case of the generalized
Poincar\'e inequality that will be stated more generally in 
Lemma \ref{lem2} in Section~\ref{becsect}.

\begin{lemma}[\textbf{Generalized Poincar{\'e} inequality, special case}]
\label{lem2b}
Let $\Lambda\subset \R^3$ be a cube of side length $L$, and define the
average of a function $f\in L^1(\Lambda)$ by $$ \langle
f\rangle_\Lambda=\frac 1{L^3} \int_\Lambda f(\x)\, d\x \ .$$ There
exists a constant $C$ such that for all measurable sets
$\Omega\subset\Lambda$ and all $f\in H^1(\Lambda)$ the inequality
\begin{equation} \label{poinchom}
 \int_{\Lambda} |f(\x)-\langle f\rangle_\Lambda |^2 d\x \leq C  
\left(L^2\int_\Omega |\nabla f(\x)|^2 d\x
+|\Omega^c|^{2/3}\int_\Lambda |\nabla f(\x)|^2 d\x \right)
\end{equation}
holds. Here $\Omega^c=\Lambda\setminus\Omega$, and 
$|\cdot|$ denotes the measure of a set.
.
\end{lemma}

\begin{proof} By scaling, it suffices to consider the case $L=1$. Using
the usual Poincar\'e-Sobolev inequality on $\Lambda$ (see
\cite{LL01}, Thm. 8.12), we infer that there exists a $C>0$ such
that
\begin{eqnarray}\nonumber
\|f-\langle f\rangle_\Lambda\|_{L^2(\Lambda)}^2&\leq& \half C
\|\nabla f\|_{L^{6/5}(\Lambda)}^2\\ &\leq&  C\left(\|\nabla
f\|_{L^{6/5}(\Omega)}^2+\|\nabla f\|_{L^{6/5}(\Omega^c)}^2\right)\
.
\end{eqnarray}
Applying H\"older's inequality $$ \|\nabla f\|_{L^{6/5}(\Omega)}
\leq \|\nabla f\|_{L^{2}(\Omega)}|\Omega|^{1/3} $$ (and the
analogue with $\Omega$ replaced by $\Omega^c$), we see that
(\ref{poinchom}) holds.
\end{proof}

Applying this result, we are now able to finish the proof of
Theorem \ref{hombecthm}. Denote by $\langle \Psi_0
\rangle_{\Lambda,\X}$ the average of $\Psi_0(\x,\X)$ over
$\x\in\Lambda$. Using Lemma \ref{lem2b}, with $\Omega=\Omega_\X$
and $f(\x)= \Psi_0(\x,\X)-\langle \Psi_0 \rangle_{\Lambda,\X}$, we
conclude that
\begin{eqnarray}\nonumber
&& \int d\X \int d\x  \left[\Psi_0(\x,\X)-\langle
\Psi_0\rangle_{\Lambda,\X}\right]^2
\\ \nonumber && \leq C\int d\X\left[L^2\int_{\Omega_\X}
|\nabla_{\x} \Psi_0(\x,\X)|^2 d\x\right. \\ &&\left.
\qquad\quad\qquad + \left(\frac{4\pi}3\right)^{2/3}
N^{2/3-8/51} \int_\Lambda |\nabla_{\x} \Psi_0(\x,\X)|^2 d\x
\right], \label{21b}
\end{eqnarray}
where we used that $|\Omega_\X^c|\leq (4\pi/3) N^{1-4/17}$. The
first integral on the right side of (\ref{21b}) tends to zero as
$N\to\infty$ by (\ref{47}), and the second term vanishes in this limit 
because
of (\ref{45}). Moreover,
\begin{eqnarray}\nonumber
&&\int\int d\X d\x  \left[\Psi_0(\x,\X)-\langle
\Psi_0\rangle_{\Lambda,\X}\right]^2\\ &&=\int\int d\X d\x
|\Psi_0(\x,\X)|^2-\frac 1N \frac 1{L^3} \int\int \gamma(\x,\, \y)
d\x d\y\ .
\end{eqnarray}
We conclude that
\begin{equation}
\liminf_{N\to\infty} \frac 1N \frac 1{L^3} \int\int \gamma(\x,\,
\y) d\x d\y \geq \lim_{N\to\infty}\int \int d\X d\x
|\Psi_0(\x,\X)|^2=1\ ,
\end{equation}
and Theorem \ref{hombecthm} is proven.
\end{proof}

As stated, Theorem \ref{hombecthm} is concerned with a simultaneous
thermodynamic and $a\to 0$ limit, where, as $N\to \infty$, the box
length $L$ is proportional to $N^{1/3}$ and $a\sim N^{-2/3}$. By
scaling, the above result is equivalent to considering a Bose gas in a
{\it fixed box} of side length $L=1$, and keeping $Na$ fixed as
$N\to\infty$, i.e., $a\sim 1/N$. The ground state energy of the
system is then, asymptotically, $N\times 4\pi Na$, and Theorem
\ref{hombecthm} implies that the one-particle reduced density
matrix $\gamma$ of the ground state converges, after division by $N$,  
to the projection
onto the constant function. An analogous result holds true for
inhomogeneous systems. This was recently shown in \cite{LS02} and
will be presented in Section \ref{becsect}.

\section{Gross-Pitaevskii Equation for Trapped Bosons} \label{sectgp}

In the recent experiments on Bose condensation (see, e.g.,
\cite{TRAP}), the particles are confined at very low temperatures 
in a `trap' where the particle density is {\em inhomogeneous}, 
contrary to the case of a large `box', where the 
density is essentially uniform.
We model the trap by a slowly varying confining
potential $ V$, with $V(\x)\to \infty $ as $|\x|\to \infty$.  
The
Hamiltonian becomes
\begin{equation}\label{trapham}
H =  \sum_{i=1}^{N}\left\{ -\mu \Delta_i +V(\x_i)\right\} +
 \sum_{1 \leq i < j \leq N} v(|\x_i - \x_j|) \ .
\end{equation}
Shifting the energy scale if necessary 
we can assume that $V$ is nonnegative.
The ground state energy, $\hbar\omega$, of
$- \mu \Delta + V(\x)$ is
a natural energy unit and the corresponding 
length unit, $\sqrt{\hbar/(m\omega)}=\sqrt{2\mu/(\hbar\omega)}
\equiv L_{\rm osc}$, is a measure of the extension 
of the trap. 

In the sequel we shall be considering a limit
where $a/L_{\rm osc}$ tends to zero while $N\to\infty$.  Experimentally
$a/L_{\rm osc}$ can be changed in two ways:  One can either vary $L_{\rm
osc}$ or $a$.  The first alternative is usually simpler in practice but
very recently a direct tuning of the scattering length itself has also
been shown to be feasible \cite{Cornish}.  Mathematically, both
alternatives
are equivalent, of course.  The first corresponds to writing
$V(\x)=L_{\rm osc}^{-2}  V_1(\x/L_{\rm osc})$ and keeping $V_1$ and
$v$ fixed.  The second corresponds to writing the interaction potential 
as $v(|\x|)=a^{-2}v_1(|\x|/a)$ like 
in \eqref{v1}, where $v_1$ has unit scattering length, 
and keeping $V$ and $v_1$ fixed. This is equivalent to the first, 
since for given $V_1$ and $v_1$ the ground state energy of (\ref{trapham}),
measured 
in units of $\hbar\omega$, depends only on $N$ and $a/L_{\rm osc}$. 
In the dilute limit when $a$ is much smaller than the mean particle 
distance, the energy becomes independent of $v_1$.

We choose $L_{\rm osc}$ as a length unit. The energy unit is
$\hbar\omega=2\mu L_{\rm osc}^{-2}=2\mu$.  Moreover, we find it
convenient to regard $V$ and $v_1$ as fixed. This justifies the notion
$E_0(N,a)$ for the quantum mechanical ground state energy.

The idea is now to use the information about the thermodynamic limiting
energy of the dilute Bose gas in a box to find the ground state
energy of  (\ref{trapham}) in an appropriate limit. This has been done
in \cite{LSY1999, LSY2d} and in this section we 
give an account of this work.
As we saw in Sections \ref{sect3d} and
\ref{sect2d} there is a difference in the $\rho$ dependence between
two and three dimensions, so we can expect a related difference now.
We discuss 3D first.

\subsection{Three Dimensions}

Associated with the quantum mechanical ground state energy problem
is the Gross-Pitaevskii (GP) energy functional \cite{G1961,G1963,P1961}
\begin{equation}\label{gpfunc3d}
\E^{\rm
GP}[\phi]=\int_{\R^3}\left(\mu|\nabla\phi|^2+V|\phi|^2+4\pi \mu
a|\phi|^4\right)d\x
\end{equation}
with the subsidiary condition \begin{equation}\label{norm}
\int_{\R^3}|\phi|^2=N.\end{equation} 
As before, $a>0$ is the scattering length of $v$. 
The corresponding energy is
\begin{equation}\label{gpen3d}
E^{\rm GP}(N,a)=\inf_{\int|\phi|^2=N}\E^{\rm GP}[\phi]= \E^{\rm
GP}[\phi^{\rm GP}],\end{equation} with a unique, positive $\phi^{\rm
GP}$. The existence of the minimizer $\phi^{\rm GP}$ is proved by
standard techniques and it can be shown to be continuously
differentiable, see \cite{LSY1999}, Sect.~2 and Appendix A. The
minimizer depends on $N$ and $a$, of course, and when this is
important we denote it by $\phi^{\rm GP}_{N,a}$.

The variational equation satisfied by the minimizer is the
{\it GP equation}
\begin{equation}\label{gpeq}
-\mu\Delta\phi^{\rm GP}(\x)+ V(\x)\phi^{\rm GP}(\x)+8\pi\mu a \phi^{\rm
GP}(\x)^3 = \mu^{\rm GP} \phi^{\rm GP}(\x),
\end{equation}
where $\mu^{\rm GP}$ is the chemical potential, given by
\begin{equation}\label{mugp}
\mu^{\rm GP}=dE^{\rm GP}(N,a)/dN=E^{\rm GP}(N,a)/N+ 
(4\pi \mu a/N)\int |\phi^{\rm GP}(\x)|^4 d\x.       
        \end{equation}

The GP theory has the following scaling property:
\begin{equation}\label{scalen}
E^{\rm GP}(N,a)=N E^{\rm GP}(1,Na),
\end{equation}
and
\begin{equation}\label{scalphi}
\phi^{\rm GP}_{N,a}(\x)= N^{1/2} \phi^{\rm GP}_{1,Na}(\x).
\end{equation}
Hence we see that the relevant parameter in GP theory is the
combination $Na$.

We now turn to the relation of $E^{\rm GP}$ and 
$\phi^{\rm GP}$ to the quantum mechanical ground state.  
If $v=0$, then the ground state of \eqref{trapham} is 
$$\Psi_{0}(\x_{1},\dots,\x_{N})=\hbox{$\prod_{i=1}^{N}$}\phi_{0}(\x_{i})
$$
with $\phi_{0}$ the normalized ground state of $-\mu \Delta + V(\x)$.
In this case 
clearly $\phi^{\rm GP}=\sqrt{N}\ \phi_{0}$, and then
$E^{\rm GP}=N\hbar\omega = E_0$.  In the other extreme, if $V(\x)=0$ for
$\x$ inside a large box of volume $L^3$ and $V(\x)= \infty$ otherwise,
then $\phi^{\rm GP} \approx \sqrt{N/L^3}$ and we get $E^{\rm
GP}(N,a) = 4\pi \mu a N^2/L^3$, which is the previously considered energy 
$E_0$ for the homogeneous gas in the low
density regime. (In this case, the gradient term in $\E^{\rm GP}$
plays no role.)

In general, we expect that for {\it dilute} gases in a suitable limit
\begin{equation}\label{approx}E_0
        \approx E^{\rm GP}\quad{\rm and}\quad \rho^{\rm QM}(\x)\approx
\left|\phi^{\rm GP}(\x)\right|^2\equiv \rho^{\rm GP}(\x),\end{equation}
where the quantum mechanical particle density in the ground state is
defined by \begin{equation} \rho^{\rm
QM}(\x)=N\int|\Psi_{0}(\x,\x_{2},\dots,\x_{N})|^2d\x_{2}\cdots
d\x_{N}.  \end{equation} {\it Dilute} means here that
\begin{equation}\bar\rho a^3\ll 1,\end{equation} where
        \begin{equation}\label{rhobar}
\bar\rho=\frac 1N\int|\rho^{\rm GP}(\x)|^2 d\x
\end{equation}
is the {\it mean density}.

The limit in which \eqref{approx} can be expected to be true 
should be chosen so that {\it all three} terms in
$\E^{\rm GP}$ make a contribution. The scaling relations \eqref{scalen} and \eqref{scalphi}
indicate that fixing
$Na$ as $N\to\infty$ is the right thing to do (and this is quite relevant since
experimentally $N$ can be quite large, $10^6$ and more,  and 
$Na$ can range from about 1 to $10^4$ \cite{DGPS}).
 Fixing $Na$ 
(which we refer to as the GP
case) also means that
we really are dealing with a dilute limit, because the mean density 
$\bar \rho$ is then of the order $N$ (since $\bar\rho_{N,a}=N\bar\rho_{1,Na}$) and hence 
\begin{equation}
a^3\bar\rho\sim N^{-2}.
\end{equation}

The precise statement of \eqref{approx} is:

\begin{theorem}[\textbf{GP limit of the QM ground state energy and
density}]\label{thmgp3}  If $N\to\infty$ with $Na$ fixed, then
\begin{equation}\label{econv}
\lim_{N\to\infty}\frac{{E_0(N,a)}}{ {E^{\rm GP}(N,a)}}=1,
\end{equation}
and
\begin{equation}\label{dconv}
\lim_{N\to\infty}\frac{1}{ N}\rho^{\rm QM}_{N,a}(\x)= \left
|{\phi^{\rm GP}_{1,Na}}(\x)\right|^2
\end{equation}
in the weak $L^1$-sense.
\end{theorem}

To describe situations where $Na$ is very large, it is appropriate 
to consider a limit where, as $N\to\infty$,  $a\gg
N^{-1}$, i.e. $Na\to\infty$, but still $\bar\rho a^3\to 0$.
 In
this case, the gradient term in the GP functional becomes
negligible compared to the other terms and the so-called {\it
Thomas-Fermi (TF) functional}
\begin{equation}\label{gtf}
\E^{\rm TF}[\rho]=\int_{\R^3}\left(V\rho+4\pi \mu a\rho^2\right)d\x
\end{equation}
arises. (Note that this functional has nothing to do with the
fermionic theory invented by Thomas and Fermi in 1927, except 
for a certain formal analogy.) It is
defined for nonnegative functions $\rho$ on $\R^3$. Its ground
state energy $E^{\rm TF}$ and density $\rho^{\rm TF}$ are defined
analogously to the GP case. (The TF functional is especially
relevant for the two-dimensional Bose gas. There  $a$ has to
decrease exponentially with $N$ in the GP limit, so the TF limit
is more adequate; see Subsection \ref{sub2d} below).

Our second main result of this section is that minimization of
(\ref{gtf}) reproduces correctly the ground state energy and
density of the many-body Hamiltonian in the limit when
$N\to\infty$, $a^3\bar \rho\to 0$, but $Na\to \infty$ (which we
refer to as the TF case), provided the external potential is
reasonably well behaved. We will assume that $V$ is asymptotically
equal to some function $W$ that is homogeneous of some order $s>0$, i.e., 
$W(\lambda\x)=\lambda^s W(\x)$ for all $\lambda>0$, 
and locally H\"older continuous (see \cite{LSY2d} for a precise
definition). This condition can be relaxed, but it seems adequate
for most practical applications and simplifies things
considerably.

\begin{theorem}[\textbf{TF limit of the QM ground state energy
and   density}]\label{thm2} Assume that $V$ satisfies 
the conditions
stated above. If $\g\equiv Na\to\infty$ as $N\to\infty$, but still
$a^3\bar\rho\to 0$, then
\begin{equation}\label{econftf}
\lim_{N\to\infty}\frac{E_0(N,a)} {E^{\rm TF}(N,a)}=1,
\end{equation}
and
\begin{equation}\label{dconvtf}
\lim_{N\to\infty}\frac{\g^{3/(s+3)}}{N}\rho^{\rm
QM}_{N,a}(\g^{1/(s+3)}\x)= \tilde\rho^{\rm TF}_{1,1}(\x)
\end{equation}
in the weak $L^1$-sense, where $\tilde\rho^{\rm TF}_{1,1}$ is the
minimizer of the TF functional under the condition $\int\rho=1$,
$a=1$, and with $V$ replaced by $W$.
\end{theorem}


In the following, we will present the essentials of the proofs 
Theorems \ref{thmgp3} and \ref{thm2}.
We will derive appropriate upper and lower bounds on the ground
state energy $E_0$. 

The proof of the lower bound in Theorem \ref{thmgp3} 
presented here is a modified version of (and partly simpler than)
the original proof in \cite{LSY1999}.

The convergence of the densities follows from
the convergence of the energies in the usual way by variation with
respect to the external potential. For simplicity, we set $\mu\equiv 1$ in
the following.

\begin{proof}[Proof of Theorems \ref{thmgp3} and \ref{thm2}] {\it Part 1: 
Upper bound to the QM energy.} To derive an upper bound on $E_0$ we
use a generalization of a trial wave function of Dyson \cite{dyson},
who used this function to give an upper bound on the ground state
energy of the homogeneous hard core Bose gas (c.f.\ Section
\ref{upsec}). It is of the form
\begin{equation}\label{ansatz}
\Psi(\x_{1},\dots,\x_{N})
    =\prod_{i=1}^N\phi^{\rm
GP}(\x_{i})F(\x_{1},\dots,\x_{N}),
\end{equation}
where $F$ is constructed in the following way:
\begin{equation}F(\x_1,\dots,\x_N)=\prod_{i=1}^N
f(t_i(\x_1,\dots,\x_i)),\end{equation} where $t_i =
\min\{|\x_i-\x_j|, 1\leq j\leq i-1\}$ is the distance of $\x_{i}$
to its {\it nearest neighbor} among the points
$\x_1,\dots,\x_{i-1}$, and $f$ is a  function of $r\geq 0$. As in 
\eqref{deff} we
choose it to be
\begin{equation}
f(r)=\left\{\begin{array}{cl} f_{0}(r)/f_0(b) \quad
&\mbox{for}\quad r<b\\ 1 &\mbox{for}\quad r\geq b,
\end{array}\right.
\end{equation}
where $f_0$ is the solution of the zero energy scattering equation
(\ref{3dscatteq}) and $b$ is some cut-off parameter of order
$b\sim \bar\rho^{-1/3}$. The function (\ref{ansatz}) is not
totally symmetric, but for an upper bound it is nevertheless an
acceptable test wave function since the bosonic ground state
energy is equal to the {\it absolute} ground state energy.

The result of a somewhat lengthy computation (see \cite{LSY1999} for details) 
is the upper bound
\begin{equation}\label{ubd}
E_0(N,a)\leq E^{\rm GP}(N,a) \left( 1+O(a\bar\rho^{1/3})\right).
\end{equation}

\bigskip\noindent {\it Part 2: Lower bound to the QM energy, GP case.}
To obtain a lower bound for the QM ground state energy the
strategy is to divide space into boxes and use the estimate on the
homogeneous gas, given in Theorem \ref{lbthm2}, in each box with
{\it Neumann} boundary conditions. One then minimizes over all
possible divisions of the particles among the different boxes.
This gives a lower bound to the energy because discontinuous wave
functions for the quadratic form defined by the Hamiltonian are
now allowed. We can neglect interactions among particles in
different boxes because $v\geq 0$. Finally, one lets the box size
tend to zero. However, it is not possible to simply approximate
$V$ by a constant potential in each box. To see this consider the
case of noninteracting particles, i.e., $v=0$ and hence $a=0$.
Here $E_0=N\hbar\omega$, but a `naive' box method gives only $\min_\x
V(\x)$ as lower bound, since it clearly pays to put all the
particles with a constant wave function in the box with the lowest
value of $V$.

For this reason we start by separating out the GP wave function in
each variable and write a general wave function $\Psi$ as
\begin{equation}\label{5.23}
\Psi(\x_{1},\dots,\x_{N})=\prod_{i=1}^N\phi^{\rm
GP}(\x_{i})F(\x_{1},\dots,\x_{N}).
\end{equation}
Here $\phi^{\rm GP}=\phi^{\rm GP}_{N,a}$ is normalized so that 
$\int|\phi^{\rm GP}|^2=N$.
Eq.\ \eqref{5.23} defines $F$ for a given $\Psi$ because $\phi^{\rm GP}$ is
everywhere strictly positive, being the ground state of the
operator $- \Delta + V+8\pi a|\phi^{\rm GP}|^2$. We now compute
the expectation value of $H$ in the state $\Psi$. Using partial
integration and the variational equation (\ref{gpeq}) for
$\phi^{\rm GP}$, we see that 
\begin{equation}\label{ener2}
\frac{\langle\Psi|H\Psi\rangle}{\langle\Psi|\Psi\rangle}-E^{\rm GP}(N,a)=4\pi a
\int |\rho^{\rm GP}|^2 +Q(F),
\end{equation}
with
\begin{equation}\label{ener3}
Q(F)=\sum_{i=1}^{N} \frac{\int\prod_{k=1}^{N}\rho^{\rm GP}(\x_k)
\left(|\nabla_i F|^2+\half\sum_{j\neq i} v(|\x_i-\x_j|)|F|^2-8\pi a
\rho^{\rm GP}(\x_i)|F|^2\right)} {\int\prod_{k=1}^{N}\rho^{\rm
GP}(\x_k)|F|^2}.
\end{equation}
We recall that $\rho^{\rm GP}(\x)=|\phi^{\rm
GP}_{N,a}(\x)|^2$. For computing the ground state energy
of $H$ we have to minimize the normalized quadratic form $Q$.
 Compared to the expression for the energy involving
$\Psi$ itself we have thus obtained the replacements
\begin{equation}\label{repl}
V(\x)\to -8\pi a\rho^{\rm GP}(\x) \quad\mbox{and}\quad
\prod_{i=1}^Nd\x_i \to \prod_{i=1}^N\rho^{\rm GP}(\x_{i})d\x_{i}\ .
\end{equation}
We now use the box method on {\it this} problem. More precisely,
labeling the boxes by an index $\alpha$, we have
\begin{equation}\label{5.26}
\inf_F Q(F)\geq \inf_{\{n_\al\}} \sum_\al \inf_{F_\al}Q_\al
(F_\al),
\end{equation}
where $Q_\al$ is defined by the same formula as $Q$  but with the
integrations limited to the box $\alpha$,  $F_{\alpha}$ is a wave
function with particle number $n_\alpha$, and the infimum is taken
over all distributions of the particles with $\sum n_\al=N$.

We now fix some $M>0$, that will eventually tend to $\infty$, and
restrict ourselves to boxes inside a cube $\Lambda_M$ of side length
$M$. Since $v\geq 0$ the contribution to
\eqref{5.26} of boxes outside this cube is easily estimated from below by 
$-8\pi Na \sup_{\x\notin \Lambda_M}\rho^{\rm GP}(\x)$, which, divided
by $N$, is arbitrarily small for $M$ large, since $Na$ is fixed and
$\pgp/N^{1/2}=\pgp_{1,Na}$ decreases faster than exponentially at
infinity (\cite{LSY1999}, Lemma A.5).

For the boxes inside the cube $\Lambda_M$ we want to use Lemma
\ref{dysonl} and therefore we must approximate $\rho^{\rm GP}$ by
constants in each box. Let $\rmax$ and $\rmin$, respectively, denote
the maximal and minimal values of $\rho^{\rm GP}$ in box $\al$. Define
\begin{equation}
\Psi_\alpha(\x_1,\dots, \x_{n_\al})=F_\alpha(\x_1,\dots, \x_{n_\al}) 
\prod_{k=1}^{n_\al}\phi^{\rm GP}(\x_k),
\end{equation}
and
\begin{equation}
\Psi^{(i)}_\alpha(\x_1,\dots, \x_{n_\al})=F_\alpha(\x_1,\dots, \x_{n_\al}) 
\prod_{\substack{k=1 \\ k\neq 
i}}^{n_\al}\phi^{\rm GP}(\x_k).
\end{equation}
We have, for all $1\leq i\leq n_\al$,
\begin{equation}\label{5.29}
\begin{split}
&\int\prod_{k=1}^{n_\alpha}\rho^{\rm GP}(\x_k)\left(|\nabla_i 
F_\alpha|^2\right.+\half\sum_{j\neq i}\left.
v(|\x_i-\x_j|)|F_\alpha|^2\right) 
\\&\geq
\rmin\int \left(|\nabla_i\Psi^{(i)}_\alpha|^2\right. +\half\sum_{j\neq i}\left.
v(|\x_i-\x_j|)|\Psi^{(i)}_\alpha|^2\right).
\end{split}
\end{equation}
We now use 
Lemma \ref{dysonl} to get, for all $0\leq \eps\leq 1$,
\begin{equation}\label{5.30}
(\ref{5.29})\geq \rmin\int \left(\varepsilon |\nabla_i\Psi^{(i)}_\alpha|^2 
+a(1-\eps)U(t_i)|\Psi^{(i)}_\alpha|^2\right)
\end{equation}
where $t_i$ is the distance to the nearest neighbor of $\x_i$, c.f., 
\eqref{2.29}, and $U$ the potential \eqref{softened}.

Since $\Psi_\alpha=\pgp(\x_i)\Psi^{(i)}_\al$ we can estimate
\begin{equation}\label{tpsial}
|\nabla_i\Psi_\alpha|^2 \leq 2\rmax |\nabla_i\Psi^{(i)}_\alpha|^2 +
2|\Psi^{(i)}_\alpha|^2 N C_M
\end{equation}
with
\begin{equation}
C_M=\frac1N\sup_{\x\in\Lambda_M}|\nabla\pgp(\x)|^2=\sup_{\x\in\Lambda_M}
|\nabla\pgp_{1,Na}(\x)|^2.  \end{equation} Since $Na$ is fixed, $C_M$
is independent of $N$. Inserting \eqref{tpsial} into
\eqref{5.30}, summing over $i$ and using $\rho^{\rm GP}(\x_i)\leq \rmax$ in 
the 
last term of \eqref{ener3} (in the box $\al$), we get
\begin{equation}\label{qalfal}
Q_\al(F_\alpha)\geq \frac{\rmin}{\rmax}E^{U}_\eps(n_\al,L)-8\pi a\rmax n_\al -
\eps C_M n_\al,
\end{equation}
where $L$ is the side length of the box and $E^{U}_\eps(n_\al,L)$ is
the ground state energy of
\begin{equation}\label{eueps}
\sum_{i=1}^{n_\al}(-\half\eps \Delta_i+(1-\eps)aU(t_i)) 
\end{equation}
in the box (c.f.\ \eqref{halfway}). We want to minimize \eqref{qalfal}
with respect to $n_\al$ and drop the subsidiary condition
$\sum_\al{n_\al}=N$ in \eqref{5.26}. This can only lower the minimum.
For the time being we also ignore the last term in
\eqref{qalfal}. (The total contribution of this term for all boxes is
bounded by $\eps C_M N$ and will be shown to be negligible compared to
the other terms.)

Since the lower bound for the energy of Theorem
\ref{lbthm2} was obtained precisely from a lower bound to the operator
\eqref{eueps}, we can use the statement and proof of Theorem \ref{lbthm2}. 
{F}rom this we see that
\begin{equation}\label{basicx}
E^{U}_\eps(n_\al,L)\geq (1-\varepsilon)\frac{4\pi
an_\al^2}{L^3}(1-CY_\al^{1/17})
\end{equation} 
with $Y_\al=a^3n_\al/L^3$, provided  $Y_\al$ is small enough, 
$\eps\geq Y_\al^{1/17}$ and
$n_\al\geq {\rm (const.)} Y_\al^{-1/17}$. The condition on $\eps$ is
certainly fulfilled if we choose $\eps=Y^{1/17}$ with
$Y=a^3N/L^3$. We now want to show that the $n_\alpha$ minimizing 
the right side of \eqref{qalfal} is large enough for \eqref{basicx} to apply.

If the minimum of the right side of \eqref{qalfal}
(without the last term) is taken for some $\bar n_\al$, we have
\begin{equation}\label{minnal}
\frac{\rmin}{\rmax}
\left(E^{U}_\eps(\bar n_\al+1,L)-E^{U}_\eps(\bar n_\al,L)\right)\geq 
8\pi a\rmax.
\end{equation}
On the other hand, we  claim that
\begin{lemma} For any $n$ 
\begin{equation}\label{chempot}
E^{U}_\eps( n+1,L)-E^{U}_\eps(n,L)\leq 8\pi a\frac{ 
n}{L^3}.
\end{equation}
\end{lemma}
\begin{proof}
Denote the operator \eqref{eueps} by $\tilde H_n$,  with $n_\alpha=n$, and 
let $\tilde 
\Psi_n$ 
be its ground state. Let $t_i'$ be the distance to the nearest neighbor
of $\x_i$ among the $n+1$ points $\x_1,\dots,\x_{n+1}$ (without $\x_i$) and $t_i$ the
corresponding distance excluding $\x_{n+1}$. Obviously, for $1\leq i\leq n$,
\begin{equation}
U(t_i')\leq U(t_i)+U(|\x_i-\x_{n+1}|)
\end{equation}
and
\begin{equation}
U(t_{n+1}')\leq \sum_{i=1}^nU(|\x_i-\x_{n+1}|).
\end{equation}
Therefore
\begin{equation}
\tilde H_{n+1}\leq \tilde H_{n}-\half\eps\Delta_{n+1}+2a\sum_{i=1}^nU(|\x_i-\x_{n+1}|).
\end{equation}
Using $\tilde\Psi_n/L^{3/2}$  as trial function for $\tilde H_{n+1}$ we arrive at
\eqref{chempot}.
\end{proof}
Eq.\ \eqref{chempot} together with \eqref{minnal} shows that
 $\bar n_\al$ is at least $\sim \rmax L^3$.
We shall choose 
$L\sim N^{-1/10}$, so
the conditions needed for (\ref{basicx}) are fulfilled for $N$ large enough, 
since $\rmax\sim N$ and hence
$\bar n_\al\sim N^{7/10}$ and
$Y_\al\sim N^{-2}$.

In order to obtain a lower bound on $Q_\al$ we therefore have to minimize
\begin{equation}\label{qalpha}
 4\pi 
a\left(\frac{\rmin}{\rmax}\frac{n_\al^2}{L^3}\left(1-CY^{1/17}\right)
-2n_\al\rmax\right).
\end{equation}
We can drop the 
requirement that $n_\al$ has to 
be an integer. The minimum of (\ref{qalpha}) is obtained for
\begin{equation}
n_\al= \frac{\rmax^2}{\rmin}\frac{L^3}{(1-CY^{1/17})}.
\end{equation}
By Eq.\ (\ref{ener2}) this gives the following lower bound, 
including now the last term in \eqref{qalfal} as well as the 
contributions from the 
boxes outside $\Lambda_M$, 
\begin{equation}\label{almostthere}
\begin{split}
&E_0(N,a)-E^{\rm GP}(N,a)\geq \\
&4\pi a\int|\rho^{\rm GP}|^2-4\pi a\sum_{\al\subset\Lambda_M} 
\rmin^2
L^3\left(\frac{\rmax^3}{\rmin^3}\frac{1}{(1-CY^{1/17})}\right)\\ &-Y^{1/17}NC_M-4\pi 
aN\sup_{\x\notin\Lambda_M}\rho^{\rm GP}(\x).
\end{split}
\end{equation}
Now $\rho^{\rm GP}$ is differentiable and strictly 
positive. Since all the boxes are in the fixed cube $\Lambda_M$ there are 
constants 
$C'<\infty$, $C''>0$,
such that
\begin{equation}
\rmax-\rmin\leq NC'L,\quad \rmin\geq NC''.
\end{equation}
Since $L\sim N^{-1/10}$ and $Y\sim N^{-17/10}$ we therefore have, for large $N$,
\begin{equation}
\frac{\rmax^3}{\rmin^3}\frac{1}{(1-CY^{1/17})}\leq 
1+{\rm (const.)}N^{-1/10}
\end{equation}
Also,
\begin{equation}
4\pi a\sum_{\al\subset\Lambda_M} \rmin^2 L^3\leq 4\pi a\int 
|\rho^{\rm GP}|^2\leq E^{\rm 
GP}(N,a).
\end{equation}
Hence, noting that $ E^{\rm GP}(N,a)=N E^{\rm GP}(1,Na)\sim N$ since $Na$ is fixed,
\begin{equation}\label{there}
\frac{E_0(N,a)}{E^{\rm GP}(N,a)}\geq 1-{\rm (const.)}(1+C_M)N^{-1/10}-{\rm (const.)}
\sup_{\x\notin \Lambda_M}|\pgp_{1,Na}|^2,
\end{equation}
where the constants depend on $Na$. We can now take $N\to\infty$ and 
then $M\to\infty$.

\bigskip
\noindent {\it Part 3: Lower bound to the QM energy, TF case.}
In the above proof of the lower bound in the GP case we did not attempt to
keep track of the dependence of the constants on $Na$. In the TF case
$Na\to\infty$, so one would need to take a closer look at this
dependence if one wanted to carry the proof directly over to this
case. But we don't have to do so, because there is a simpler direct
proof. Using the explicit form of the TF minimizer, namely
\begin{equation}\label{tfminim}
\rho^{\rm TF}_{N,a}(\x)=\frac 1{8\pi a}[\mu^{\rm TF}-V(\x)]_+,
\end{equation}
where  $[t]_+\equiv\max\{t,0\}$ and $\mu^{\rm TF}$ is chosen so
that the normalization condition $\int \rho^{\rm TF}_{N,a}=N$
holds, we can use 
\begin{equation}\label{vbound}
V(\x)\geq \mu^{\rm TF}-8\pi a \rho^{\rm
TF}(\x) 
\end{equation}
to get a replacement as in (\ref{repl}), but without
changing the measure. Moreover, $\rho^{\rm TF}$ has compact
support, so, applying again the box method described above, the
boxes far out do not contribute to the energy. However, $\mu^{\rm
TF}$ (which depends only on the combination $Na$) tends to
infinity as $Na\to\infty$. We need to control the
asymptotic behavior of $\mtf$, and this leads to
the restrictions on $V$ described in the paragraph preceding
Theorem \ref{thm2}. For simplicity, we shall here only consider the case when $V$ 
itself is homogeneous, i.e., $V(\lambda\x)=\lambda^sV(\x)$ for all $\lambda>0$ with 
some $s>0$.  

In the same way as in \eqref{mugp} we have, with $g=Na$, 
\begin{equation}\label{mutf}
\mu^{\rm TF}(g)=dE^{\rm TF}(N,a)/dN=E^{\rm TF}(1,g)+ 
4\pi g\int |\rho^{\rm TF}_{1,g}(\x)|^2 d\x.     
        \end{equation}
The TF energy, chemical potential and minimizer 
satisfy the scaling relations
\begin{equation}
E^{\rm TF}(1,g)=g^{s/(s+3)}E^{\rm TF}(1,1),
\end{equation}
\begin{equation}
\mu^{\rm TF}(g)=g^{s/(s+3)} \mu^{\rm TF}(1)  ,
\end{equation}
and
\begin{equation}
g^{3/(s+3)}\rho^{\rm TF}_{1,g}(g^{1/(s+3)}\x)= \rho^{\rm TF}_{1,g}(\x)  .
\end{equation}
We also introduce the scaled interaction potential, $\widehat v$, by
\begin{equation}
\widehat v(\x)  =g^{2/(s+3)}v(g^{1/(s+3)}\x)
\end{equation}
with scattering length 
\begin{equation}
\widehat a=g^{-1/(s+3)}a.
\end{equation}
 Using \eqref{vbound}, \eqref{mutf} and 
the scaling relations we obtain
\begin{equation}
E_0(N,a)\geq E^{\rm TF}(N,a)+4\pi N g^{s/(s+3)}\int |\rho^{\rm TF}_{1,1}|^2 +
g^{-2/(s+3)}Q
\end{equation}
with
\begin{equation}
Q=\inf_{\int|\Psi|^2=1}\sum_{i}\int\left(|\nabla_i\Psi|^2\right.+\half \sum_{j\neq i}
\left.\widehat v(\x_i-\x_j)|\Psi|^2-8\pi
N\widehat a \rtf_{1,1}(\x_i)|\Psi|^2\right).
\end{equation}
We can now proceed exactly as in Part 2 to arrive at the the analogy of 
Eq.\ \eqref{almostthere}, which in the present case becomes
\begin{equation}\label{almosttherex}
\begin{split}
&E_0(N,a)-E^{\rm TF}(N,a)\geq \\
&4\pi N g^{s/(s+3)}\int |\rho^{\rm TF}_{1,1}|^2-4\pi N\widehat a\sum_{\al} 
\rmax^2
L^3(1-C\widehat Y^{1/17})^{-1}.
\end{split}
\end{equation}
Here $\rmax$ is the maximum of $\rho^{\rm TF}_{1,1}$ in the box
$\alpha$, and $\widehat Y=\widehat a^3 N/L^3$. This holds as long as $L$ does not
decrease too fast with $N$. In particular, if $L$ is simply fixed, this
holds for all large enough $N$. Note that
\begin{equation}
\bar\rho=N\bar\rho_{1,g}\sim N g^{-3/(s+3)} \bar\rho_{1,1},
\end{equation}
so that $\widehat a^3 N\sim a^3 \bar
\rho$ goes to zero as $N\to\infty$ by assumption. Hence, if we first let
$N\to\infty$ (which implies $\widehat Y\to 0$) and then take $L$ to zero, we
of arrive at the desired 
result
\begin{equation}\label{lowertf}
\liminf_{N\to\infty}\frac{E_0(N,a)}{E^{\rm TF}(N,a)}\geq 1
\end{equation}
in the limit $N\to\infty$, $a^3\bar\rho\to 0$. Here
we used the fact that (because $V$, and hence $\rtf$, is continuous by
assumption) the Riemann sum $\sum_\al\rmax^2 L^3$ converges to
$\int|\rtf_{1,1}|^2$ as $L\to 0$.  Together with the upper bound (\ref{ubd}) 
and
the fact that $E^{\rm GP}(N,a)/E^{\rm TF}(N,a)=E^{\rm GP}(1,Na)/E^{\rm
TF}(1,Na)\to 1$ as $Na\to\infty$, which holds under our regularity
assumption on $V$ (c.f.\ Lemma 2.3 in \cite{LSY2d}), this proves
(\ref{econv}) and (\ref{econftf}).

\bigskip
\noindent {\it Part 4: Convergence of the densities.} The
convergence of the energies implies the convergence of the
densities in the usual way by variation of the external potential.
We show here the TF case, the GP case goes analogously. Set again
$\g=Na$. Making the replacement
\begin{equation}
V(\x)\longrightarrow V(\x)+\delta\g^{s/(s+3)}Z(\g^{-1/(s+3)}\x)
\end{equation}
for some positive $Z\in C_0^\infty$ and redoing the upper and
lower bounds we see that (\ref{econftf}) holds with $W$ replaced
by $W+\delta Z$. Differentiating with respect to $\delta$ at
$\delta=0$ yields
\begin{equation}
\lim_{N\to\infty}\frac{\g^{3/(s+3)}}N\rho^{\rm
QM}_{N,a}(\g^{1/(s+3)}\x) =\tilde\rho^{\rm TF}_{1,1}(\x)
\end{equation}
in the sense of distributions. Since the functions all have
$L^1$-norm 1, we can conclude that there is even weak
$L^1$-convergence.
\end{proof}

\subsection{Two Dimensions}\label{sub2d}

In contrast to the three-dimensional case the energy per particle for
a dilute gas in two dimensions is {\it nonlinear} in $\rho$. In view
of Schick's formula \eqref{2den} for the energy of the homogeneous gas
it would appear natural to take the interaction into account in two
dimensional GP theory by a term
\begin{equation}
4\pi\int_{\R^2} |\ln(|\phi(\x)|^2 a^2)|^{-1}|\phi(\x)|^4{
d}\x,\end{equation} 
and such a term has, indeed, been suggested in
\cite{Shev} and \cite{KoSt2000}.  However, since the nonlinearity
appears only in a logarithm, this term is unnecessarily complicated
as far as leading order computations are concerned.  For dilute gases
it turns out to be sufficient, to leading order, to use an interaction
term of the same form as in the three-dimensional case, i.e, define the 
GP functional as (for simplicity we put $\mu=1$ in this section)
\begin{equation}\label{2dgpfunc}
\E^{\rm
GP}[\phi]=\int_{\R^2}\left(|\nabla\phi|^2+V|\phi|^2+4\pi 
\alpha|\phi|^4\right)d\x,
\end{equation}
where instead of $a$ the coupling constant is now
\begin{equation}\label{alpha}\alpha=|\ln(\bar\rho_N 
a^2)|^{-1}\end{equation}
with $\bar\rho_N$ the {\em mean density}
for the GP functional 
at coupling constant 
$1$ and particle number $N$. This is defined analogously to \eqref{rhobar} 
as
\begin{equation}
\bar\rho_N=\frac1N\int|\phi^{\rm GP}_{N,1}|^4d\x
\end{equation}
where $\phi^{\rm GP}_{N,1}$ is the minimizer of \eqref{2dgpfunc} with 
$\alpha=1$ and subsidiary condition $\int|\phi|^2=N$.
Note that $\alpha$ in \eqref{alpha} depends on 
$N$ through the mean density.

Let us denote the GP energy 
for a given $N$ and coupling constant 
$\alpha$ by $E^{\rm GP}(N,\alpha)$ and the corresponding minimizer by
$\phi^{\rm GP}_{N,\alpha}$.
As in three dimensions the scaling relations
\begin{equation}E^{\rm GP}(N,\alpha)=NE^{\rm GP}(1,N\alpha)\end{equation}
and
    \begin{equation}N^{-1/2}\phi^{\rm GP}_{N,\alpha}=\phi^{\rm GP}_{1,N\alpha},
\end{equation}
hold, and the relevant parameter is
\begin{equation}g\equiv N\alpha.\end{equation}

In three dimensions, where $\alpha=a$, 
it is natural to consider the limit $N\to\infty$ with $g=Na$= const.
The analogue of Theorem \ref{thmgp3} in two dimensions is
\begin{theorem}[{\bf Two-dimensional GP limit 
    theorem}] 
\label{2dlimit}
If, for $N\to\infty$,\linebreak $a^2\brtf_N\to 0$ with
$g=N/|\ln(a^2\brtf_N)|$ fixed, then
\begin{equation}\label{econv2}
\lim_{N\to\infty}\frac{E_{0}(N,a)}{\Egp(N,1/|\ln(a^2\brtf_N)|)}=
1
\end{equation}
and
\begin{equation}\label{dconv2}
\lim_{N\to\infty}\frac{1}{ N}\rho^{\rm QM}_{N,a}(\x)= \left
|{\phi^{\rm GP}_{1,g}}(\x)\right|^2
\end{equation}
in the weak $L^1$-sense.
\end{theorem}

This result, however, is of rather limited use in practice.  The reason is
that in two dimensions the scattering length has to
decrease exponentially with $N$ if $g$ is fixed.  
The parameter $g$ is
typically {\it very large} in two dimensions 
so it is more appropriate to consider the
limit $N\to\infty$ and $g\to\infty$ (but still $\bar\rho_N a^2\to
0$).

For  potentials $V$ that are {\it homogeneous} functions of $\x$, i.e., 
\begin{equation}\label{homog}V(\lambda \x)=\lambda^sV(\x)\end{equation}
for some $s>0$, this limit can be described by the a
`Thomas-Fermi' energy functional like \eqref{gtf} with coupling constant unity:
\begin{equation}\label{tffunct}
\E^{\rm TF}[\rho]=\int_{\R^2}\left(V(\x)\rho(\x)+4\pi \rho(\x)^2\right)
{ d}\x.
\end{equation}
This is just the GP functional without the gradient term and $\alpha=1$.
Here $\rho$ is a nonnegative function on $\R^2$ and the normalization 
condition is 
\begin{equation}\label{norm2}\int\rho(\x)d\x=1.\end{equation}

The minimizer of \eqref{tffunct} can be given explicitly.  It is
\begin{equation}\label{tfminim2}\rho^{\rm 
TF}_{1,1}(\x)=(8\pi)^{-1}[\mu^{\rm TF}-V(\x)]_+\end{equation}
where the chemical potential 
$\mu^{\rm TF}$ is determined by the normalization condition \eqref{norm2} 
and $[t]_{+}=t$ 
for $t\geq 0$ and zero otherwise.
We denote the corresponding energy by $E^{\rm TF}(1,1)$.
By scaling one obtains
\begin{equation}\lim_{g\to\infty} 
    E^{\rm GP}(1,g)/g^{s/(s+2)}=E^{\rm TF}(1,1),\end{equation}
 \begin{equation}\label{gptotf}\lim_{g\to\infty}g^{2/(s+2)}
\rho^{\rm 
GP}_{1,g}(g^{1/(s+2)}\x)=\rho^{\rm TF}_{1,1}(\x),\end{equation}
with the latter limit in the  strong $L^2$ sense.

Our main result about two-dimensional 
Bose gases in  external potentials satisfying \eqref{homog}  
is that analogous limits also hold for the many-particle quantum mechanical 
ground state at
low densities:
\begin{theorem}[{\bf Two-dimensional TF limit theorem}]\label{thm22}
In two dimensions, if
$a^2\bar\rho_N\to 0$, but $g=N/|\ln(\bar\rho_N a^2)|\to \infty$ as 
$N\to\infty$ 
then
\begin{equation}\lim_{N\to \infty}\frac{E_0(N,a)}{g^{s/s+2}}= 
E^{\rm TF}(1,1)\end{equation}
and, in the weak $L^1$ sense,
\begin{equation}\label{conv}\lim_{N\to\infty}\frac{g^{2/(s+2)}}N
\rho^{\rm 
QM}_{N,a}(g^{1/(s+2)}\x)=\rho^{\rm TF}_{1,1}(\x).\end{equation}
\end{theorem}

\noindent {\it Remarks:} 1. As in Theorem \ref{thm2}, it is sufficient that $V$ is
asymptotically equal to some homogeneous potential, $W$. In this case,
$E^{\rm TF}(1,1)$ and $\rho^{\rm TF}_{1,1}$ in Theorem \ref{thm22}
should be replaced by the corresponding quantities for $W$.

2. From Eq.\ \eqref{gptotf} it follows that
\begin{equation}\bar\rho_N\sim N^{s/(s+2)}\end{equation} for large $N$.
Hence the low density criterion 
$a^2\bar\rho\ll 1$, means that
$a/L_{\rm osc}\ll  N^{-s/2(s+2)}$.

We shall now comment briefly on the proofs of Theorems 
\ref{2dlimit} and \ref{thm22}, 
mainly pointing out the differences from the 3D case considered previously.

The upper bounds for the energy are obtained exactly in a same way as
in three dimensions. For the lower bound in Theorem \ref{2dlimit} the
point to notice is that the expression
 \eqref{qalpha}, that has to be minimized over $n_\al$, is in 2D
 replaced by
\begin{equation}\label{qalpha2}
 4\pi 
\left(\frac{\rmin}{\rmax}\frac{n_\al^2}{L^2}\frac1{|\ln(a^2n_\alpha/L^2)|}
\left(1-\frac C{|\ln(a^2N/L^2)|^{1/5}}\right)
-\frac{2n_\al\rmax}{|\ln(a^2\bar\rho_N)|}\right),
\end{equation}
since Eq.\ \eqref{basicx} has to be 
replaced by the analogous inequality for 2D (c.f.\ \eqref{lower}).
To minimize \eqref{qalpha2} we use the following lemma:
\begin{lemma}\label{xb}
For $0<x,b<1$ and $k\geq 1$ we have
\begin{equation}
\frac{x^2}{|\ln x|}-2\frac b{|\ln b|}xk\geq -
\frac{b^2}{|\ln b|}\left(1+\frac 1{(2|\ln b|)^2}\right)k^2.
\end{equation}
\end{lemma}
\begin{proof} Replacing $x$ by $x/k$ and using the monotonicity of $\ln$ we 
see that it suffices to consider $k=1$.
Since $\ln x\geq-\frac 1{de}x^{-d}$ for
all $d>0$ we have
\begin{equation}
\frac{x^2}{b^2}\frac{|\ln b|}{|\ln x|}
-2\frac xb\geq\frac{|\ln b|}{b^2}ed x^{2+d}-\frac{2x}{b}
\geq c(d)(b^ded\,|\ln b|)^{-1/(1+d)}
\end{equation}
with
\begin{equation}
c(d)=2^{(2+d)/(1+d)}\left(\frac 1{(2+d)^{(2+d)/(1+d)}}-\frac 1
{(2+d)^{1/(1+d)}}\right)\geq -1-\frac 14d^2.
\end{equation}
Choosing $d=1/|\ln b|$ gives the desired result.
\end{proof}

Applying this lemma with $x=a^2n_\al/L^2$, $b=a^2\rmax$ and 
\begin{equation}k=\frac{\rmax}{\rmin}\,
\left(1-\frac C{|\ln(a^2N/L^2)|^{1/5}}\right)^{-1}\frac{|\ln(a^2\rmax)|}
{|\ln(a^2\bar\rho_N)|}
\end{equation}
we get the bound
\begin{equation}
\eqref{qalpha2}\geq -4\pi\frac{\rmax^2L^2}{|\ln(a^2\bar\rho_N)|}
\left(1+\frac1{4|\ln(a^2\rmax)|^2}\right) k.
\end{equation}
In the limit considered, $k$ and the factor in parenthesis both tend to 1 and 
the Riemann sum over the boxes $\alpha$ converges to the integral as $L\to 0$.

The TF case, Thm.\ \ref{thm22}, is treated in the same way as in three 
dimensions, with modifications analogous to those just discussed when passing 
from 3D to 2D in GP theory.

\bigskip
\section{BEC for Dilute Trapped Gases}\label{becsect}

It is gratifying to see the experimental realization, in traps, of
the long-predicted Bose-Einstein condensation (BEC) of gases. From
the theoretical point of view, however, a rigorous demonstration
of this phenomenon -- starting from the many-body Hamiltonian of
interacting particles -- has not yet been achieved. Following
\cite{LS02}, we will provide in this section  such a rigorous
justification for the ground state of 2D or 3D bosons in a trap with
repulsive pair potentials, and in the well-defined limit in which the
Gross-Pitaevskii (GP) formula is applicable. It is the first proof of
BEC for interacting particles in a continuum (as distinct from
lattice) model and in a physically realistic situation. The
Gross-Pitaevskii limit under discussion here is, of course, a
physically simpler limit than the usual thermodynamic limit in which
the average density is held fixed as the particle number goes to
infinity. In the GP limit one also lets the range of the potential go
to zero as $N$ goes to infinity, but in such a way that the overall
effect is non-trivial. That is, the combined effect of the infinite
particle limit and the zero range limit is such as to leave a
measurable residue --- the GP function.

It was shown in the previous section (see also Theorem
\ref{compthm} below) that, for each  fixed $Na$, the minimization
of the GP functional correctly reproduces the large $N$
asymptotics of the ground state energy and density of $H$ -- but
no assertion about BEC in this limit was made. We will now extend
this result by showing that in the Gross-Pitaevskii limit  there
is indeed 100\% Bose condensation in the ground state. This is a
generalization of the homogeneous case considered in Theorem
\ref{hombecthm}. In the following, we concentrate on the 3D case,
but analogous considerations apply also  to the 2D case.

For use later, we define the projector
\begin{equation}
P^{\rm GP}= |\varphi^{\rm GP}\rangle\langle \varphi^{\rm GP}|\ .
\end{equation}
Here (and everywhere else in this section) we denote $\varphi^{\rm
GP}=\phi^{\rm GP}_{1,Na}$ for simplicity, where $\phi^{\rm
GP}_{1,Na}$ is the minimizer of the GP functional (\ref{gpfunc3d})
with parameter $Na$ and normalization condition $\int|\phi|^2=1$
(compare with (\ref{scalphi})). Moreover, we set $\mu\equiv 1$.

In the following, $\Psi_0$ denotes the (nonnegative and normalized)
ground state of the Hamiltonian (\ref{trapham}). BEC refers to the
reduced one-particle density matrix $ \gamma(\x,\x')$ of $\Psi_0$,
defined in (\ref{defgamma}).

Complete (or 100\%) BEC is defined to be the property that
$\mbox{$\frac{1}{N}$}\gamma(\x,\x')$ not only has an eigenvalue of
order one, as in the general case of an incomplete BEC, but in the
limit it has only one nonzero eigenvalue (namely 1). Thus,
$\mbox{$\frac{1}{N}$}\gamma(\x,\x')$ becomes a simple product
$\varphi(\x)^*\varphi(\x')$ as $N\to \infty$, in which case $\varphi$ is called
the {\it condensate wave function}.  In the GP limit, i.e.,
$N\to\infty$ with $N a$ fixed, we can show that this is the case, and
the condensate wave function is, in fact, the GP minimizer $\varphi^{\rm GP}$.

\begin{theorem}[\textbf{Bose-Einstein condensation}]\label{becthm}
For each fixed $Na$ $$ \lim_{N\to\infty} \frac 1 N \gamma(\x, \x')
= \varphi^{\rm GP}(\x)\varphi^{\rm GP}(\x')\ . $$ in trace norm, i.e., $\Tr \left|\frac 1
N \gamma - P^{\rm GP} \right| \to 0$.
\end{theorem}

We remark that Theorem \ref{becthm} implies that there is also 100\%
condensation for all $n$-particle reduced density matrices 
\begin{eqnarray}\nonumber
&&\gamma^{(n)}(\x_1,\dots,\x_n;\x_1',\dots,\x_n')\\&&=n!\binom{N}{n}\int
\Psi_0(\x_1,\dots,\x_N)\Psi_0(\x_1',\dots,\x_n',\x_{n+1},
\dots\x_N)d\x_{n+1}\cdots d\x_N\nonumber \\
\end{eqnarray}
of
$\Psi_0$, i.e., they converge, after division by the normalization factor,  
to the one-dimensional projector onto
the $n$-fold tensor product of $\varphi^{\rm GP}$. In other words, for 
$n$ fixed particles the probability of finding them all in the same state 
$\varphi^{\rm GP}$ tends to 1 in the 
limit considered. To see this,
let $a^*, a$ denote the boson creation and annihilation operators
for the state $\varphi^{\rm GP}$, and observe that
\begin{equation}
1\geq \lim_{N\to\infty} N^{-n}\langle \Psi_0 | (a^*)^n a^n|\Psi_0\rangle =
\lim_{N\to\infty} N^{-n} \langle \Psi_0 | (a^*a)^n|\Psi_0\rangle  \  ,
\end{equation}
since the terms coming from the commutators $[a, a^*]=1$ are of
lower order as $N\to \infty$ and vanish in the limit. From
convexity it follows that
\begin{equation}
N^{-n}  \langle \Psi_0 | (a^*a)^n|\Psi_0\rangle \geq N^{-n} \langle
\Psi_0 | a^*a|\Psi_0\rangle ^n \,
\end{equation}
which converges to $1$ as $N\to\infty$, proving our claim.

Another corollary, important for the interpretation of
experiments, concerns the momentum distribution of the ground
state.

\begin{corollary}[\textbf{Convergence of momentum distribution}] Let
$$\widehat\rho (\k)=\int \int\gamma(\x, \x') \exp [i \k\cdot (\x
-\x')]
 d\x d\x'$$
denote the one-particle momentum  density of $\Psi_0$. Then, for
fixed $Na$, $$ \lim_{N\to\infty} \frac 1N
\widehat\rho(\k)=|\widehat\varphi^{\rm GP}(\k)|^2 $$ strongly in
$L^1(\R^3)$. Here, $\widehat\varphi^{\rm GP}$ denotes the Fourier
transform of $\varphi^{\rm GP}$.
\end{corollary}

\begin{proof} If ${\mathcal F}$ denotes the (unitary) operator `Fourier
transform' and if $h$ is an arbitrary $L^\infty$-function,
then
\begin{eqnarray}\nonumber
\left|\frac 1N\int \widehat\rho h-\int |\widehat\varphi^{\rm
GP}|^2 h\right|&=&\left|\Tr[{\mathcal F}^{-1}
(\gamma/N-P^{\rm GP}){\mathcal F}h]\right|\\ \nonumber
&\leq& \|h\|_\infty \Tr |\gamma/N-P^{\rm GP}|,
\end{eqnarray}
from which we conclude that $$\|\widehat\rho/N-|\widehat\varphi^{\rm
GP}|^2 \|_1\leq \Tr|\gamma/N-P^{\rm GP}|\ .$$
\end{proof}

Before proving Theorem \ref{becthm}, let us state some prior
results on which we shall build. Then we shall formulate two
lemmas, which will allow us to prove Theorem \ref{becthm}.

The following theorem is an extension of Theorem \ref{thmgp3}.

\begin{theorem}[\textbf{Asymptotics of the energy components}]\label{compthm}
If $\psi_0$ denotes \linebreak the solution to the zero-energy scattering equation for
$v$ (under the boundary condition
$\lim_{|\x|\to\infty}\psi_0(\x)=1$) and
$s=\int|\nabla\psi_0|^2/(4\pi a)$, then $0<s\leq 1$ and
\begin{subequations}\label{parttwo}
\begin{eqnarray}\nonumber
&&\lim_{N\to\infty} \int  |\nabla_{\x_1} \Psi_0(\x_1,\X)|^2 d\x_1\,
d\X
\\ \label{3a}
 &&\qquad= \int|\nabla\varphi^{\rm GP}(\x)|^2d\x + 4\pi Na s \int|\varphi^{\rm GP}(\x)|^4
d\x,\\  &&\lim_{N\to\infty} \int  V(\x_1)|\Psi_0(\x_1,\X)|^2 d\x_1\,
d\X = \int V(\x) |\varphi^{\rm GP}(\x)|^2 d\x, \\ \nonumber
&&\lim_{N\to\infty} \half\sum_{j=2}^N \int
v(|\x_1-\x_j|)|\Psi_0(\x_1,\X)|^2 d\x_1\, d\X
\\ \label{part2} &&\qquad=(1-s) 4\pi Na \int|\varphi^{\rm GP}(\x)|^4 d\x.
\end{eqnarray}
\end{subequations}
\end{theorem}
Here we introduced again the short hand notation
\begin{equation}
\X=(\x_2,\dots,\x_N)\qquad{\rm and}\quad  d\X=\prod\limits_{j=
2}^N d\x_j.
\end{equation}

Theorem \ref{compthm} is a simple consequence of Theorem
\ref{thmgp3} by variation with respect to the different
components, in the same way as was explained in the proof of Theorem
\ref{hombecthm}, c.f.\ Eqs.\ \eqref{45}--\eqref{dmua}. This was also
noted in \cite{CS01a}.

As already stated, Theorem \ref{becthm} is a generalization of Theorem
\ref{hombecthm}, the latter corresponding to the case that $V$ is a
box potential. It should be noted, however, that we use different
scaling conventions in these two theorems: In Theorem \ref{hombecthm}
the box size grows as $N^{1/3}$ to keep the density fixed, while in
Theorem \ref{becthm} we choose to  keep the confining external
potential fixed. Both conventions are equivalent, of course, c.f.\ the
remark at the end of Section \ref{sectbe}, but when comparing the
exponents of $N$ that appear in the proofs of the two theorems the
different conventions should be born in mind.

There are two essential components of our proof of Theorem
\ref{becthm}. The first is a proof that the part of the kinetic
energy that is associated with the interaction $v$ (namely, the
second term in (\ref{3a})) is mostly located in small balls
surrounding each particle. More precisely, these balls can be
taken to have radius roughly $N^{-5/9}$, which is much smaller
than the mean-particle spacing $N^{-1/3}$. This allows us to
conclude  that the function of $\x$ defined for each fixed value
of $\X$ by
\begin{equation}\label{defff}
f_\X(\x)=\frac 1{\varphi^{\rm GP}(\x)} \Psi_0(\x,\X)\geq 0
\end{equation}
has the property that $\nabla_\x f_\X(\x)$ is almost zero outside
the small balls centered at points of $\X$.

The complement of the small balls has a large volume but it can be
a weird set; it need not even be connected. Therefore, the
smallness of $\nabla_\x f_\X(\x)$ in this set does not guarantee
that $f_\X(\x)$ is nearly constant (in $\x$), or even that it is
continuous. We need $f_\X(\x)$ to be nearly constant in order to
conclude BEC. What saves the day is the knowledge that the total
kinetic energy of $f_\X(\x)$ (including the balls) is not huge.
The result that allows us to combine these two pieces of
information in order to deduce the almost constancy of $f_\X(\x)$
is the generalized Poincar\'e inequality in Lemma \ref{lem2}.

Using the results of Theorem \ref{compthm}, partial integration
and the GP equation (i.e., the variational equation for $\varphi^{\rm GP}$,
see Eq. (\ref{gpeq})) we see that
\begin{equation}\label{bound}
\lim_{N\to\infty} \int  |\varphi^{\rm GP}(\x)|^2 |\nabla_\x f_\X(\x)|^2 d\x\,
d\X
 = 4\pi Na s\int |\varphi^{\rm GP}(\x)|^4 d\x\ .
\end{equation}
The following Lemma shows that to leading order all the energy in
(\ref{bound}) is concentrated in small balls.

\begin{lemma}[\textbf{Localization of the energy}]\label{lem1}
For fixed $\X$ let
\begin{equation}\label{defomega} \Omega_\X=\left\{\x\in \R^3
\left| \, \min_{k\geq 2}|\x-\x_k|\geq
N^{-1/3-\delta}\right\}\right.
\end{equation} for some $0<\delta< 2/9$. Then $$ \lim_{N\to\infty}
\int d\X \int_{\Omega_\X} d\x |\varphi^{\rm GP}(\x)|^2 |\nabla_\x f_\X(\x)|^2
= 0\ . $$
\end{lemma}

\noindent {\it Remark.} In the proof 
of Theorem \ref{hombecthm} we chose $\delta$ to be 4/51, but the
following proof shows that one can extend the range of $\delta$ beyond
this value.
\begin{proof}
We shall show that
\begin{eqnarray} \nonumber &&\int
d\X \int_{\Omega_\X^c} d\x\, |\varphi^{\rm GP}(\x)|^2 |\nabla_\x f_\X(\x)|^2\\
\nonumber &&+\half\int d\X \int d\x\, |\varphi^{\rm GP}(\x)|^2 \sum_{k\geq 2}
v(|\x-\x_k|) |f_\X(\x)|^2\\ \nonumber &&- 8\pi Na \int d\X \int
d\x\, |\varphi^{\rm GP}(\x)|^4 |f_\X(\x)|^2\\ \label{lowbound}&& \geq -4\pi Na
\int|\varphi^{\rm GP}(\x)|^4 d\x - o(1)
\end{eqnarray}
as $N\to \infty$, which implies the assertion of the Lemma by
virtue of (\ref{bound}) and the results of Theorem \ref{compthm}.
Here, $\Omega_\X^c$ is the complement of $\Omega_\X$. The proof of
(\ref{lowbound}) is actually just a detailed examination of the
lower bounds to the energy derived in \cite{LSY1999} and
\cite{LY1998} and described in Sections 2 and 5.
We use the same methods as there,
just describing the differences from the case considered here.

Writing
\begin{equation}
f_\X(\x)=\prod_{k\geq 2}\varphi^{\rm GP}(\x_k)F(\x,\X)
\end{equation}
and using that $F$ is symmetric in the particle coordinates, we
see that (\ref{lowbound}) is equivalent to
\begin{equation}\label{qf}
\frac 1N Q_\delta(F)\geq -4\pi Na \int|\varphi^{\rm GP}|^4 - o(1),
\end{equation}
where $Q_\delta$ is the quadratic form
\begin{eqnarray}\nonumber Q_\delta(F)&=&\sum_{i=1}^{N} \int_{\Omega_i^c} |\nabla_i
F|^2\prod_{k=1}^{N}|\varphi^{\rm GP}(\x_k)|^2d\x_k\\ \nonumber &&+\sum_{1\leq
i<j\leq N} \int
v(|\x_i-\x_j|)|F|^2\prod_{k=1}^{N}|\varphi^{\rm GP}(\x_k)|^2d\x_k\\
\label{qf2} &&-8\pi Na\sum_{i=1}^{N} \int
|\varphi^{\rm GP}(\x_i)|^2|F|^2\prod_{k=1}^{N}|\varphi^{\rm GP}(\x_k)|^2d\x_k.
\end{eqnarray}
Here $\Omega_i^c$ denotes the set
$$\Omega_i^c=\{(\x_1,\X)\in\R^{3N}| \, \min_{k\neq
i}|\x_i-\x_k|\leq N^{-1/3-\delta}\}.$$

While  (\ref{qf}) is not true for all conceivable $F$'s satisfying
the normalization condition $$\int
|F(\x,\X)|^2\prod_{k=1}^{N}|\varphi^{\rm GP}(\x_k)|^2d\x_k=1,$$ it {\it is}
true for an $F$, such as ours, that has bounded kinetic energy
(\ref{bound}). Looking at Section 5, we see that Eqs.
\eqref{ener2}--\eqref{ener3}, \eqref{almostthere}--\eqref{there} 
are similar to (\ref{qf}),
(\ref{qf2}) and almost establish (\ref{qf}), but there are
differences which we now explain.

In our case, the kinetic energy of particle $i$ is restricted
to the subset of $\R^{3N}$ in which $\min_{k\neq i}|\x_i-\x_k|\leq
N^{-1/3-\delta}$. However, looking at the proof of the lower bound
to the ground state energy of a homogeneous Bose gas discussed in Section 2,
which enters the proof of Theorem \ref{thmgp3}, we
see that if we choose $\delta\leq 4/51$ only this part of the
kinetic energy is needed for the lower bound, except for
some part with a relative magnitude of the order
$\eps=O(N^{-2\alpha})$ with $\alpha=1/17$. (Here we use the a
priori knowledge that the kinetic energy is bounded by
\eqref{bound}. See also the
analogous discussion in Section \ref{sectbe}, p. \pageref{qffff}.) 
We can even do better and choose some
$4/51<\delta<2/9$, if $\alpha$ is chosen small enough. (To be
precise, we choose $\beta=1/3+\alpha$ and $\gamma=1/3-4\alpha$ in
the notation of  (\ref{ans}), and $\alpha$ small enough). The
choice of $\alpha$ only affects the magnitude of the error term,
however, which is still $o(1)$ as $N\to\infty$.

Proceeding exactly as in  Section 5 and taking the differences just mentioned 
into account we arrive at (\ref{qf}).
\end{proof}

In the following, $\K\subset\R^m$ denotes a bounded and connected
set that is sufficiently nice so that the Poincar\'e-Sobolev
inequality (see \cite{LL01}, Thm. 8.12) holds on $\K$. In
particular, this is the case if $\K$ satisfies the cone property
\cite{LL01} (e.g., if $\K$ is a ball or a cube). The next Lemma
generalizes Lemma \ref{lem2b}.

\begin{lemma}[\textbf{Generalized Poincar{\'e} inequality}]\label{lem2}
For $d\geq 2$ let $\K\subset\R^d$ be as explained above, and let
$h$ be a bounded function with $\int_\K h=1$. There exists a
constant $C$ (depending only on $\K$ and $h$) such that for all
measurable sets $\Omega\subset\K$ and all $f\in H^1(\K)$  with
$\int_\K f h\, d\x=0$, the inequality
\begin{equation} \label{poinc}
 \int_{\K} |f(\x)|^2 d\x \leq C \left(\int_\Omega |\nabla f(\x)|^2 d\x
+\left(\frac{|\Omega^c|}{|\K|}\right)^{2/d}\int_\K |\nabla
f(\x)|^2 d\x \right)
\end{equation}
holds. Here $|\cdot|$ denotes the measure of a set, and
$\Omega^c=\K\setminus\Omega$.
\end{lemma}

\begin{proof} By the usual Poincar\'e-Sobolev inequality on $\K$ (see
\cite{LL01}, Thm. 8.12),
\begin{eqnarray}\nonumber
\|f\|_{L^2(\K)}^2&\leq& \tilde C \|\nabla
f\|_{L^{2d/(d+2)}(\K)}^2\\ &\leq& 2\tilde C\left(\|\nabla
f\|_{L^{2d/(d+2)}(\Omega)}^2+\|\nabla
f\|_{L^{2d/(d+2)}(\Omega^c)}^2\right),
\end{eqnarray}
if $d\geq 2$
and $\int_\K f h=0$. Applying H\"older's inequality $$ \|\nabla
f\|_{L^{2d/(d+2)}(\Omega)} \leq \|\nabla
f\|_{L^{2}(\Omega)}|\Omega|^{1/d} $$ (and the analogue with
$\Omega$ replaced by $\Omega^c$), we see that (\ref{poinc}) holds
with $C=2|\K|^{2/d}\tilde C$.
\end{proof}

The important point in Lemma \ref{lem2} is  that there is no
restriction on $\Omega$ concerning regularity or connectivity.

\begin{proof}[Proof of Theorem \ref{becthm}]
For some $R>0$ let $\K=\{\x\in\R^3, |\x|\leq R\}$, and define $$
\langle f_\X\rangle_\K=\frac 1{\int_\K |\varphi^{\rm GP}(\x)|^2 d\x} \int_\K
|\varphi^{\rm GP}(\x)|^2 f_\X(\x)\, d\x \  . $$ We shall use Lemma \ref{lem2},
with $d=3$, $h(\x)=|\varphi^{\rm GP}(\x)|^2/\int_\K|\varphi^{\rm GP}|^2$,
$\Omega=\Omega_\X\cap\K$ and $f(\x)= f_\X(\x)-\langle f_\X
\rangle_\K$ (see (\ref{defomega}) and (\ref{defff})). Since $\varphi^{\rm GP}$
is bounded on $\K$ above and below by some positive constants,
this Lemma also holds (with a different constant $C'$) with $d\x$
replaced by $|\varphi^{\rm GP}(\x)|^2d\x$ in (\ref{poinc}). Therefore,
\begin{eqnarray}\nonumber
&& \int d\X \int_\K d\x |\varphi^{\rm GP}(\x)|^2 \left[f_\X(\x)-\langle
f_\X\rangle_\K\right]^2
\\ \nonumber && \leq C'\int d\X\left[\int_{\Omega_\X\cap \K}
|\varphi^{\rm GP}(\x)|^2|\nabla_{\x} f_\X(\x)|^2 d\x\right. \\ &&\left.
\qquad\quad\qquad + \frac {N^{-2\delta}}{R^2} \int_\K
|\varphi^{\rm GP}(\x)|^2|\nabla_{\x} f_\X(\x)|^2 d\x \right], \label{21}
\end{eqnarray}
where we used that $|\Omega_\X^c\cap\K|\leq (4\pi/3)
N^{-3\delta}$. The first integral on the right side of (\ref{21})
tends to zero as $N\to\infty$ by Lemma \ref{lem1}, and the second
is bounded by (\ref{bound}). We conclude, since $$\int_\K
|\varphi^{\rm GP}(\x)|^2 f_\X(\x) d\x\leq \int_{\R^3} |\varphi^{\rm GP}(\x)|^2
f_\X(\x)d\x$$ because of the positivity of $f_\X$, that
\begin{eqnarray}\nonumber \liminf_{N\to\infty} \frac 1N \langle
\varphi^{\rm GP}|\gamma|\varphi^{\rm GP}\rangle &\geq& \int_\K |\varphi^{\rm GP}(\x)|^2 d\x \,
\lim_{N\to\infty}\int d\X \int_\K d\x |\Psi_0(\x,\X)|^2
\\ \nonumber &=&\left[\int_\K |\varphi^{\rm GP}(\x)|^2 d\x\right]^2,
\end{eqnarray}
where the last equality follows from (\ref{dconv}). Since the
radius of $\K$ was arbitrary, we conclude that
$$\lim_{N\to\infty}\frac 1 N \langle\varphi^{\rm GP}|\gamma|\varphi^{\rm GP}\rangle= 1,$$
implying convergence of $\gamma/N$ to $P^{\rm GP}$ in
Hilbert-Schmidt norm. Since the traces are equal, convergence even
holds in trace norm  (cf. \cite{S79}, Thm. 2.20), and Theorem
\ref{becthm} is proven.
\end{proof}

We remark that the method presented here also works in the case of a
two-dimensional Bose gas. The relevant parameter to be kept fixed in
the GP limit is $N/|\ln (a^2 \bar\rho_N)|$, all other considerations carry over
without essential change, using the results in \cite{LSY2d,LY2d},
c.f.\ Sections 3 and 5.2. A minor difference concerns the parameter
$s$ in Theorem \ref{compthm}, which can be shown to be always equal to
$1$ in 2D, i.e., the interaction energy is purely kinetic in the GP
limit (see \cite{CS01b}). It should be noted that the existence of BEC
in the ground state in 2D is not in conflict with its absence at
positive temperatures \cite{Ho,M}. In the hard core lattice gas at
half filling precisely this phenomenon occurs \cite{KLS}. We also
point out that our method necessarily fails for the one-dimensional
Bose gas, where there is presumably no BEC \cite{PiSt}. An analogue of
Lemma \ref{lem1} cannot hold in the 1D case since even a hard core
potential with arbitrarily small range produces an interaction energy
that is not localized on scales smaller than the mean particle spacing.

\bigskip

\section{The Charged Bose Gas}

The setting now changes abruptly. Instead of particles interacting
with a short-range potential $v(|\x_i-\x_j|)$ they interact via
the Coulomb potential $$v(|\x_i-\x_j|)  = |\x_i-\x_j|^{-1} $$ (in
3 dimensions). There are $N$ particles in a large box $\Lambda$ of volume
$L^3$ as before, with $\rho =N/L^{3}$.

To offset the huge Coulomb
repulsion (which would drive the particles to the walls of the box)
we add a uniform negative background of precisely the same charge,
namely density $\rho$. Our Hamiltonian is thus
\begin{equation}\label{foldyham}
H=  \sum_{i=1}^{N} - \mu \Delta_i -V(\x_i) +
 \sum_{1 \leq i < j \leq N} v(|\x_i - \x_j|)  +C
\end{equation}
with $$ V(\x)=\rho \int_{\Lambda} |\x-\y |^{-1}d\y  \qquad \qquad
{\rm and}\qquad \qquad
 C= \frac{1}{ 2} \rho \int_{\Lambda} V(\x)d\x\ .
$$

Each
particle interacts only with others and not with itself. Thus, despite 
the fact that the Coulomb potential is positive definite, the
ground state energy $E_0$ can
be (and is) negative (just take $\Psi=$const.). This time, {\it
large} $\rho$ is the `weakly interacting' regime.

We know from the work in
\cite{LN} that the thermodynamic limit $e_0(\rho)$ defined as in
(\ref{eq:thmlimit}) exists.

Another way in which this problem is different from the previous one
is that {\it perturbation theory is correct to leading order}. If
one computes $(\Psi, H \Psi)$ with  $\Psi=$const, one gets the right
first order answer, namely $0$. It is the next order in $1/\rho$ that
is interesting, and this is {\it entirely} due to correlations.
In 1961 Foldy \cite{FO} calculated this correlation energy according
to the prescription of Bogolubov's 1947 theory. That theory was not
exact for the dilute Bose gas, as we have seen, even to first order. 
We are now looking at {\it second} order, which should be even
worse. Nevertheless, there was good physical intuition that this
calculation should be asymptotically {\it exact}. Indeed it is, as
proved in \cite{LS}.

The Bogolubov theory states that the main contribution to the
energy comes from pairing of particles into momenta $\k, -\k$ and
is the bosonic analogue of the BCS theory of superconductivity
which came a decade later. I.e., $\Psi_0$ is a sum of products of
terms of the form $\exp\{i\k \cdot (\x_i-\x_j)\}$.

Foldy's energy, based on Bogolubov's ansatz, has now been proved. His
calculation essentially implies an upper bound as proved by Dyson in
\cite{D2} for a slight reformulation of the model.  The lower bound is
the hard part.

\begin{theorem}[\textbf{Foldy's law}] 
\begin{equation}\label{foldyen}
\lim_{\rho\to\infty}\rho^{-1/4}e_0(\rho)\geq
-\frac{2}{5}\frac{\Gamma(3/4)}{\Gamma(5/4)}\left(\frac{2}{\mu\pi}\right)^{1/4}.
\end{equation}
\end{theorem}

This is the {\it first example} (in more than 1 dimension) in which
Bogolubov's pairing theory has been rigorously validated. It has to be
emphasized, however, that Foldy and Bogolubov rely on the existence of
Bose-Einstein condensation. We neither make such a hypothesis nor does
our result for the energy imply the existence of such condensation. As
we said earlier, it is sufficient to prove condensation in small boxes
of fixed size.

Incidentally, the one-dimensional example for which Bogolubov's
theory is asymptotically exact to the first two orders  (high
density) is the repulsive delta-function Bose gas \cite{LL}.

To appreciate the  $-\rho^{1/4}$ nature of (\ref{foldyen}), it is useful to
compare it with what one would get if the bosons had infinite mass,
i.e., the first term in (\ref{foldyham}) is dropped. Then the energy would
be proportional to $-\rho^{1/3}$ as shown in \cite{LN}. Thus, the effect
of quantum mechanics is to lower $\frac{1}{3}$ to $\frac{1}{4}$.

It is supposedly true that there is a critical mass above which the
ground state should show crystalline ordering (Wigner crystal), but this
has never been proved and it remains an intriguing open problem, even
for the infinite mass case. A simple scaling shows that 
large mass is the same as small $\rho$, and  is thus outside the
region where a Bogolubov approximation can be expected to hold.

Another important remark about the $-\rho^{1/4}$  law is its relation
to the $-N^{7/5}$ law for a $\mathit{two}$-component charged Bose gas.
Dyson \cite{D2} proved that the ground state energy for such a gas was at
least as negative as $-(\mathrm{const})N^{7/5}$  as $N\to \infty$.  Thus,
thermodynamic stability (i.e., a linear lower bound) fails for this gas.
Years later, a lower bound of this $-N^{7/5}$ form was finally established
in \cite{CLY}, thereby proving that this law is correct.  The connection
of this $-N^{7/5}$ law with the jellium $-\rho^{1/4}$ law (for which a
corresponding lower bound was also given in \cite{CLY}) was pointed out
by Dyson \cite{D2} in the following way. Assuming the correctness of
the $-\rho^{1/4}$  law, one can treat the 2-component gas by treating
each component as a background for the other. What should the density
be? If the gas has a radius $L$ and if it has $N$ bosons then $\rho =
N L^{-3}$. However, the extra kinetic energy needed to compress the
gas to this radius is $N L^{-2}$. The total energy is then $N L^{-2}
- N \rho^{1/4}$, and minimizing this with respect to $L$ leads to the
$-N^{7/5}$ law. A proof going in the other direction is in \cite{CLY}.

A problem somewhat related to bosonic jellium is \textit{fermionic} jellium.
Graf and Solovej \cite{GS} have proved that the first two terms are what
one would expect, namely
\begin{equation}
        e_{0}(\rho)=C_{\rm TF}\rho^{5/3}-C_{\rm D}\rho^{4/3}+o(\rho^{4/3}),
\end{equation}
where $C_{\rm TF}$ is the usual Thomas-Fermi constant and $C_{\rm D}$ is the
usual Dirac exchange constant.

As for the dilute Bose gas there are several relevant length scales in
the problem of the charged Bose gas. For the dilute gas there were three scales. This time
there are just two. Because of the long range nature of the Coulomb
problem there is no scale corresponding to the scattering length
$a$. One relevant length scale is again the interparticle distance
$\rho^{-1/3}$. The other is the correlation length scale 
$\ell_{\rm cor}\sim \rho^{-1/4}$ (ignoring the dependence on $\mu$). 
The order of the correlation length scale can be understood
heuristically as follows. Localizing on a scale $\ell_{\rm cor}$ requires
kinetic energy of the order of $\ell_{\rm cor}^{-2}$. The Coulomb potential
from the particles and background on the scale $\ell_{\rm cor}$ is
$(\rho\ell_{\rm cor}^3)/\ell_{\rm cor}$. Thus the kinetic energy and the Coulomb
energy balance when $\ell_{\rm cor}\sim\rho^{-1/4}$. This heuristics is
however much too simplified and hides the true complexity of the
situation. 

Note that in the high density limit 
$\ell_{\rm cor}$ is long compared to the interparticle distance. 
This is analogous to the dilute gas where 
the scale $\ell_c$ is also long compared to the interparticle
distance [see (\ref{scales})]. There is however no real analogy between the scale
$\ell_{\rm cor}$ for the charged gas and the scale $\ell_c$ for the
dilute gas. In particular, whereas $e_0(\rho)$ for the dilute gas is
up to a constant of the same order as the kinetic energy
$\sim\mu\ell_c^{-2}$ we have for the charged gas that
$e_0(\rho)\not\sim \ell_{\rm cor}^{-2}=\rho^{1/2}$. The reason for
this difference is that on average only a small fraction of the particles in the
charged gas actually correlate.

\subsection{A Short Sketch of the Rigorous Proof} 
\hfill\\

{\bf Step 1: Localization into small boxes:}
As mentioned above an important step in the rigorous proof is again to divide 
the big box $\Lambda$ into smaller boxes of some fixed size
$\ell$. This time we must require
$\ell_{\rm cor}\ll\ell$. We again use Neumann boundary conditions
on the boundary of each little box. 

In contrast to the dilute gas we can however no longer simply ignore
the interaction between the small boxes. 
To overcome this problem we use a sliding technique first introduced in 
\cite{CLY}. To explain this technique we introduce the localized interaction
$w(\x,\y)=\chi(\x)Y_\omega(\x-\y)\chi(\y)$, where
$Y_\omega(\x)=|\x|^{-1}\exp(-\omega |\x|)$ is a Yukawa potential and $\chi$ 
is a smooth approximation to the characteristic function of the cube 
$[0,\ell]^3$. Consider the Hamiltonian
\begin{eqnarray}
  H_\ell&=&\sum_{j=1}^N\left(-\mu\gamma^{-1}\Delta_{\ell,j}
    -\rho\int w(\x_j,\y)\,d\y \right)\nonumber
  \\&&{}
  +\sum_{1\leq i<j\leq N}w(\x_i,\x_j)
  +\mfr1/2\rho^2\iint
  w(\x,\y)\,d\x\,d\y,\label{eq:Hnell}
\end{eqnarray}
where $-\Delta_\ell$ is the Neumann Laplacian for
the cube $[0,\ell]^3$, and $\gamma$ is a constant depending on $\chi$
that converges to 1 as
$\chi$ converges to the characteristic function of $[0,\ell]^3$. 
We consider $-\Delta_\ell$ as acting in the Hilbert space $L^2(\R^3)$
in such a way that it is zero on functions supported away from
$[0,\ell]^3$. 
Let $H_\ell(\z)$ be the corresponding
Hamiltonian for the translated box $\z+[0,\ell]^3$. If we ignore small
errors (that can be fairly easily controlled) due to the boundary of
$\Lambda$, the original operator  
$H$ in (\ref{foldyham}) is bounded below by the operator
\begin{equation}\label{eq:sliding}
  \ell^{-3}\gamma \int_{\Lambda}H_\ell(\z) d\z -\frac{\omega N}{2\ell}.  
\end{equation}
The last error term is due to the interaction between the small
boxes. This requires that the parameter $\omega$ is chosen
appropriately. In fact, it must tend to infinity as $\chi$ converges to the
characteristic function of $[0,\ell]^3$. If we make sure that 
$\ell\gg \omega\rho^{-1/4}=\omega\ell_{\rm cor}$
we have that $\omega N/\ell\ll N\rho^{1/4}$ and we may ignore this
error term.  
 
In the treatment of the dilute gas it was important to understand the
distribution of particles into the different cells. This is not an
issue here. In fact the presence of the background will imply that the
smallest energy is achieved when the number of particles in each 
cell essentially neutralizes the background.  More precisely we may
simply look for a lower bound  on the energy independent of
the numbers of particles in each little box. 

The operator $H_\ell$ was
defined as an operator on $L^2(\R^{3N})$, but we may
consider its restriction to the invariant subspace with precisely $n$
particles in the box $[0,\ell]^3$. This restriction is equivalent to
an operator $H^n_\ell$ acting on the space 
$L^2([0,\ell]^{3n})$.  Note that  $H^n_\ell$ has the form 
(\ref{eq:Hnell}) with $N$ replaced by $n$, except that now the Neumann 
Laplacian is considered as acting in $L^2([0,\ell]^3)$. 

The problem of finding a lower bound on the ground state energy of $H$
has been reduced to finding a lower bound on the operator $H^n_\ell$
independently of $n$ for $0\leq n\leq N$. Without discussing the proof 
of this point further we shall from now on use
that it is essentially enough to consider $n=\rho\ell^3$ (the neutral
case). 
 
{\bf Step 2: Reducing to quadratic Hamiltonian:} The next step is to use the
second quantization 
formalism, which is the
one used by Bogolubov and which is a very convenient bookkeeping device
(but it has to be noted that it is no more than a convenient device and
it does not introduce any new physics or mathematics).
In second quantized form we 
may write the Hamiltonian $H^n_\ell$ as
\begin{eqnarray}
  H^n_{\ell}&=&
  \mu\gamma^{-1}\sum_\p|\p|^2 a^*_\p\an_\p
  +\mfr{1}/{2}\sum_{\p\q,\vecmu\vecnu}\hw_{\p\q,\vecmu\vecnu}
  a^*_\p a^*_\q\an_\vecnu \an_\vecmu
  -\rho\ell^3\sum_{\p\q}\hw_{\0\p,\0\q}a^*_\p\an_\q\nonumber\\&&
  +\mfr{1}/{2}\rho^2\ell^6\hw_{\0\0,\0\0},
\end{eqnarray}
where $a^*_\p$ is the creation operator for the eigenfunction $u_\p$ 
of the Neumann Laplacian and 
\begin{equation}
\hw_{\p\q,\vecmu\vecnu}=\iint w(\x,\y)u_\p(\x)u_\q(\y)u_\vecmu(\x)u_\vecnu(\y)
\,d\x\,d\y.
\end{equation}

Through a very complicated bootstrapping procedure one now proves that
to the order of interest here one may ignore several terms in
$H^n_\ell$ and consider instead the Hamiltonian
\begin{eqnarray}
  H_{\rm Q}
  &=&\mu\gamma^{-1}\sum_\p|\p|^2 a^*_\p\an_\p\nonumber\\&&{}
  +\sum_{\p\q\ne\0}\hw_{\p\q,\0\0}\left(
    a^*_\p\an_\q a^*_\0\an_\0 +
    \mfr{1}/{2}a^*_\p a^*_\q\an_\0 \an_\0+\mfr{1}/{2}a^*_\0 a^*_\0\an_\p
    \an_\q\right).
\end{eqnarray}
We have denoted this operator with a Q since it is quadratic in the
operators $a^*_\p$, with $\p\ne\0$. 

In order to reduce to this quadratic Hamiltonian it is important to be 
able to control the degree of condensation into the Neumann ground
state, the constant function  $u_\0=\ell^{-3/2}$. More precisely, if we denote by
$\hn_+=\sum_{\p\ne\0}a^*_\p\an_\p$ the operator counting the number of
particles not in the condensate, we would like to know that for the
minimal energy state, the expectation $\langle \hn_+\rangle$ is small
compared to the total particle number $n$. This, it turns out is not
too difficult. One needs, however, also
a good bound on $\langle \hn_+^2\rangle$ and this is more difficult.
In \cite{LS} this is not achieved directly through a bound on $\langle
\hn_+^2\rangle$ in the ground state. 
Rather it is proved that one may change the ground 
state without changing its energy very much,
so that it only contains values of $\hn_+$ localized close to 
$\langle \hn_+\rangle$. This technique, which in \cite{LS} was called
``localization of large matrices'', allows one to consider only states 
where $\langle \hn_+^2\rangle\approx\langle \hn_+\rangle^2$.

{\bf Step 3: Controlling the kinetic energy:} The final reduction of the
Hamiltonian before one can apply the Bogolubov-Foldy method concerns
the kinetic energy. In fact, if $\Delta$ denotes the Laplacian on $\R^3$
we shall use the bound (see Lemma~6.1 in \cite{LS}) 
\begin{equation}
\langle\phi,-\Delta_\ell\phi\rangle\geq
\left\langle\phi,\upchi_\ell F(-\Delta)\upchi_\ell\phi\right\rangle,
\hbox{ where}\quad F(v)=(1-Ct)\frac{v^2}{v+(\ell
  t^3)^{-2}},
\end{equation}
for functions $\phi$ orthogonal to constants and where we have assumed 
that the parameter $t$ is chosen such that 
$\|\partial^\alpha\chi\|_\infty\leq t^{-|\alpha|}\ell^{-|\alpha|}$
for all multi-indices $\alpha$ with $|\alpha|\leq 3$. Here $C>0$ is
some universal constant. 

If we introduce the operator 
$b^*_\k=a^*(\chi_\k)\an_\0$, where $\chi_\k$ is the projection 
of the function $\chi(\x)\exp(i\k\cdot\x)$ onto the orthogonal complement of
the constant functions, we can write the above 
inequality as 
\begin{equation}
\sum_\p|\p|^2a^*_\p\an_\p\geq (2\pi)^{-3}n^{-1}\int F(|\k|^2) b^*_\k\bn_\k d\k.
\end{equation}
Note the extra factor $n^{-1}$ which is due to the fact that $b^*_\k$
contains a factor $\an_\0$ and that we are considering states with
total particle number $n$.

The reason that we have included a factor $\an_\0$ in $b^*_\k$ is in
order to be able to write the the last sum 
in $H_Q$ also in terms of the $b^*_\k$. 
In fact, if we use that $w(\x,\y)=\chi(\x)Y_\omega(\x-\y)w(\y)$ we get that 
\begin{equation}\label{eq:HQlower}
   H_{\rm Q}\geq \mfr1/2 (2\pi)^{-3}\int_{\R^3}h(\k)\,d\k-\sum_{\p\q\ne\0}\hw_{\p\q,\0\0}
a^*_\p\an_\q,
\end{equation}
where
\begin{eqnarray}
  h(\k)&=& \mu\gamma^{-1} n^{-1}F(|\k|^2) 
  \left(b_\k^*\bn_\k+b_{-\k}^*\bn_{-\k}\right)\nonumber \\&&{}
  +\widehat Y_\omega(\k)\ell^{-3}\left(b^*_\k\bn_\k+b^*_{-\k}b_{-\k}+b^*_{\k}b^*_{-\k}
      +\bn_\k\bn_{-\k}\right).
\end{eqnarray}
The last term in (\ref{eq:HQlower}) comes from the fact that one must
commute the factor $\an_\0a^*_\0$ in $b^*_\k\bn_\k$ so that it occurs in
the normal ordered way $a^*_\0\an_\0$ as in $H_Q$. It is not difficult to see
that one may ignore the last error term in (\ref{eq:HQlower}) to the order of interest here. 

{\bf Step 4. Applying Bogolubov's method:}
The last step in the analysis is now to use the following simple form of Bogolubov's
method. 

\begin{theorem}[{\bf Simple case of Bogolubov's method}]\label{thm:bogolubov}\hfill\\
For constants $\cA\geq\cB>0$ and all $\k$ we have the operator inequality 
\begin{eqnarray*}
\lefteqn{\cA(b^*_\k\bn_\k+b^*_{-\k}\bn_{-\k})+\cB(b^*_\k b^*_{-\k}+\bn_\k\bn_{-\k})}&&\\ 
&\geq&-\mfr{1}/{2}(\cA-\sqrt{\cA^2-\cB^2})
([\bn_{\k},b^*_{\k}]+[\bn_{-\k},b^*_{-\k}]).
\end{eqnarray*}
\end{theorem}
\begin{proof}
We may complete the square
\begin{eqnarray*}
\lefteqn{\cA(b^*_\k\bn_\k+b^*_{-\k}\bn_{-\k})+\cB(b^*_\k b^*_{-\k}+\bn_\k\bn_{-\k})}&&\\
&=&D(b^*_\k+\alpha\bn_{-\k})(\bn_\k+\alpha b^*_{-\k})
+D(b^*_{-\k}+\alpha\bn_{\k})(\bn_{-\k}+\alpha b^*_{-\k})\\
&&-D\alpha^2([\bn_{\k},b^*_{\k}]+[\bn_{-\k},b^*_{-\k}]),
\end{eqnarray*}
where
$$
D(1+\alpha^2)=\cA,\quad2D\alpha=\cB.
$$
We choose the solution 
$
\alpha={\cA}/{\cB}-\sqrt{{\cA^2}/{\cB^2}-1}.
$
Hence
$$
D\alpha^2=\cB\alpha/2=\mfr{1}/{2}(\cA-\sqrt{\cA^2-\cB^2}).
$$
\end{proof}

It is not difficult to see that $[\bn_\k,b_\k^*]\leq n \ell^3$. 
Using the theorem above we see that the ground state energy of $H^n_\ell$
up to the errors we have ignored is bounded below by 
$
 -\mfr{1}/{2}(2\pi)^{-3}\int f(\k)-(f(\k)^2-g(\k)^2)^{1/2}d\k,
$
where 
$g(\k)=n\widehat{Y}_\omega(\k)$ and
$f(\k)=g(\k)+\mu\gamma^{-1}\ell^3F(|\k|^2)$. 

Up to the order of interest we may now replace  $\gamma^{-1}F(|\k|^2)$ by $|\k|^2$ and 
$Y_\omega$ by the Coulomb potential
and thus $\widehat{Y}_\omega(\k)$ by $4\pi|\k|^{-2}$. If we now also 
replace $n$ by $\rho\ell^3$ we see that the ground state energy of
$H^n_\ell$ to leading order is given by 
$$
  -\mfr{1}/{2}(2\pi)^{-3}\rho\ell^3\int (4\pi|\k|^{-2}+\rho^{-1}\mu|\k|^2
    -((4\pi|\k|^{-2}+\rho^{-1}\mu|\k|^2)^2
    -(4\pi|\k|^{-2})^2)^{1/2}d\k
$$
$$=2^{-1/2}\pi^{-3/4}\rho\ell^3(\rho/\mu)^{1/4}
  \int_0^\infty1+x^4-x^2(2+x^4)^{1/2}dx.
$$
If we finally use that the integral above is 
$$
\frac{2^{3/4}\sqrt{\pi}\Gamma(3/4)}{5\Gamma(5/4)}
$$
and that the leading order of the ground state energy of the original operator $H$ according
to (\ref{eq:sliding}) is $L^3/\ell^3$ times the ground state energy of
$H^n_\ell$ (again ignoring $\gamma$) 
we arrive at Foldy's law (\ref{foldyen}).

\bibliographystyle{amsalpha}

\bigskip\medskip

\end{document}